\documentclass[3p,times,twocolumn,sort&compress]{elsarticle}

\usepackage{ecrc-ac}
\usepackage[british]{babel}
\usepackage{color}

\usepackage{amsmath} 

\volume{00}

\firstpage{1}

\journalname{Nuclear Physics B Proceedings Supplement}

\runauth{L.N. Mihaila, M. Steinhauser}

\jid{nuphbp}

\jnltitlelogo{Nuclear Physics B Proceedings Supplement}

\usepackage{amssymb}

\usepackage[figuresright]{rotating}

\begin{document}

\begin{frontmatter}



\dochead{}

\title{
  Selected topics on multi-loop calculations to Higgs boson properties and
  renormalization group functions}


\author{Luminita N. Mihaila and Matthias Steinhauser}

\address{Institut f{\"u}r Theoretische Teilchenphysik, Karlsruhe
  Institute of Technology (KIT), 76128 Karlsruhe, Germany}

\begin{abstract}
  We review some results obtained in the context of the Collaborative Research
  Center/Transregio~9. In particular we discuss three-loop corrections to the
  Higgs boson mass in the Minimal Supersymmetric Standard Model, higher order
  corrections to Higgs boson production, and
  the calculations of renormalization group functions and decoupling
  constants.
\end{abstract}



\end{frontmatter}

\graphicspath{ {figs_c5/} }

\renewcommand{\textfraction}{0}
\renewcommand{\topfraction}{1}
\renewcommand{\bottomfraction}{1}

\newcommand{\msbar}{$\overline{\text{MS}}$}
\newcommand{\drbar}{$\overline{\text{DR}}$}
\newcommand{\msbarformel}{\overline{\text{MS}}}

\newcommand{\parform}{{\tt ParFORM}}
\newcommand{\form}{{\tt FORM}}
\newcommand{\tform}{{\tt TFORM}}

\newcommand{\drbarmod}{$\overline{\mbox{\abbrev MDR}}$}
\newcommand{\mususy}{\mu_{\rm SUSY}}
\newcommand{\mugut}{\mu_{\rm GUT}}

\newcommand{\gsim}{\;\rlap{\lower 3.5 pt \hbox{$\mathchar \sim$}} \raise 1pt
 \hbox {$>$}\;}
\newcommand{\lsim}{\;\rlap{\lower 3.5 pt \hbox{$\mathchar \sim$}} \raise 1pt
 \hbox {$<$}\;}

\sloppy




\section{\label{sec::intro}Introduction}

The discovery of a Higgs boson in July 2012 at the LHC has given a big boost
to particle physics, both to theoretical developments and to experimental
analyses. On the experimental side an intensive study of the properties of the
newly discovered particle has started. Among them are measurements of the
mass, the spin, the couplings to fermions and bosons, the self-coupling and
the decay width. At the same time the search for ``New Physics'', i.e.,
phenomena that are not described by the Standard Model (SM) of particle
physics, have been intensified. The experimental efforts are supported by
theoretical studies which concentrate on the one hand on the development of
new theories which can be tested experimentally. On the other hand, higher
order quantum corrections are computed which are necessary to match the
precision reached by the experimental collaborations. In this review we
discuss several calculations in the context of the
Higgs boson which have been performed 
within the Collaborative Research Center/Transregio~9 (CRC/TR~9).
Some of the calculations are performed within the framework of the SM, others
in extensions like the Minimal Supersymmetric Standard Model (MSSM).

Within the MSSM there are five physical Higgs bosons, two CP-even, one CP-odd
and a charged one.  A central feature of the MSSM is the prediction of the
mass of the lightest CP-even Higgs boson. At lowest order it is bounded
by the $Z$ boson mass but higher order corrections, which are significant, can
raise the value to above 130~GeV. Thus, there are parameter sets of the MSSM
which are consistent with the Higgs boson mass value of about 125~GeV as
measured at the LHC. In Section~\ref{sec:mh} we describe the code {\tt H3m}
which is the only publicly available program containing complete
strong three-loop
corrections.

A crucial input in the experimental analyses concerned with Higgs boson
properties are precise predictions of its production cross section.
Section~\ref{sec::ggh} discusses several calculations in this context.  Among
them are the next-to-next-to-leading order (NNLO) QCD corrections performed in
full theory. Taking the limit of large top quark mass allows a quantitative
check of the effective-theory result which is implemented in most of the
computer codes.  Furthermore, we summarize the first steps towards the
third-order corrections which are needed for the phenomenological analyses
since sizeable corrections are expected. Moreover, this calculation is
quite challenging from the technical point of view.  Section~\ref{sec::ggh}
also contains a summary of supersymmetric (SUSY) corrections to Higgs boson
production in the gluon fusion channel, in particular, three-loop corrections
to the matching coefficient of the effective Higgs-gluon coupling.

Finally, Section~\ref{sec::beta} is devoted to renormalization group functions
and decoupling relations. A particular emphasis is put on the recent
calculation of the three-loop corrections to the beta functions of the
SM couplings. Furthermore, we describe the decoupling procedure
which is relevant when crossing flavour thresholds within the SM
but also for the transition from the SM to the MSSM and from the
MSSM to a Grand Unified Theory (GUT). As applications we discuss
low-energy theorems which relate decoupling constants with effective Higgs
boson couplings, gauge coupling unification in the minimal SUSY SU(5) model,
and the stability of the Higgs potential in the SM.



\section{\label{sec:mh}Higgs boson mass in the MSSM}

The Higgs boson mass measurement by ATLAS 
$M_h=125.36\pm 0.37(stat.)\pm 0.18(syst.)$~GeV~\cite{Aad:2014aba} and CMS 
$M_h=125.02^{+0.26}_{-0.27}(stat.)^{+0.14}_{-0.15}(syst.)$~GeV~\cite{Khachatryan:2014jba} 
already reached the accuracy level of a precision
observable. All the other properties (couplings, spin and parity) of the new
particle have been determined with significantly lower accuracy. They are
consistent with the SM predictions based on a minimal Higgs sector. However,
also, various other beyond-the-SM (BSM) theories with a richer Higgs sector
can accommodate the present experimental data~\cite{Aad:2014aba,Khachatryan:2014jba}.
While within the SM the Higgs boson mass is a free parameter, in BSM theories
it can often be predicted, providing an important test of the model.  For
example, the mass of the lightest Higgs boson within supersymmetric models is,
beyond the tree-level approximation, a function of the top squark masses and
mixing parameters. It grows logarithmically with the top squark masses and can
be used to determine the supersymmetric mass scale, once the mixing parameters
are fixed. This approach has received considerable attention
recently~\cite{Feng:2013tvd,Hahn:2013ria,Draper:2013oza}, partially because
the direct searches for supersymmetric particles at the LHC remained
unsuccessful, indicating a possible lower bound for the supersymmetric mass
scale in the TeV range.

In the following, we concentrate on the precise prediction of the lightest
Higgs boson mass within the MSSM.  Compared to the SM, the MSSM Higgs sector
is described by two additional parameters, usually chosen to be the
pseudo-scalar mass $M_A$ and the ratio of the vacuum expectation values of the
two Higgs doublets, $\tan\beta=v_2/v_1$. The masses of the other Higgs bosons
are then fixed by supersymmetric constraints. In particular, the mass of the
light CP-even Higgs boson, $M_h$, is bounded from above.  At tree-level, the
mass matrix of the neutral CP-even Higgs bosons $h$, $H$ has the following
form:
\begin{eqnarray}
 \lefteqn{
  {\cal M}_{H,\rm tree}^2 =
  \frac{\sin 2\beta}{2}\times
}
\\
\lefteqn{
  \left(
  \begin{array}{cc}
    M_Z^2 \cot\beta + M_A^2 \tan\beta &
    -M_Z^2-M_A^2 \\
    -M_Z^2-M_A^2 &
    M_Z^2 \tan\beta + M_A^2 \cot\beta
  \end{array}
  \right)
  \,.
}
\nonumber
\end{eqnarray}
The diagonalization of ${\cal M}_{H,\rm tree}^2$ gives the tree-level results
for $M_h$ and $M_H$, and leads to the well-known bound $M_h < M_Z$ which is
approached in the limit $\tan\beta\to \infty$.  Radiative corrections to the
Higgs pole mass raise this bound substantially.  The dominant radiative
corrections are generated by the top quark and top squark loops that scale
like $\sim \alpha_t m_t^2\sim m_t^4$ ($m_t$ is the top quark mass and
$\sqrt{\alpha_t}$ is proportional to the top-Yukawa
coupling)~\cite{Ellis:1990nz,Okada:1990vk,Haber:1990aw}.

Including higher order corrections, one obtains for the Higgs boson mass
matrix
\begin{eqnarray}
  {\cal M}_{H}^2 &=&
  {\cal M}_{H,\rm tree}^2 -
  \left(
  \begin{array}{cc}
    \hat\Sigma_{\phi_1}       & \hat\Sigma_{\phi_1\phi_2} \\
    \hat\Sigma_{\phi_1\phi_2} & \hat\Sigma_{\phi_2}
  \end{array}
  \right)
  \,,
  \label{eq::MH}
\end{eqnarray}
which again gives the physical Higgs boson masses upon diagonalization.
The renormalized quantities $\hat\Sigma_{\phi_1}$, $\hat\Sigma_{\phi_2}$
and $\hat\Sigma_{\phi_1\phi_2}$ are obtained from the self energies of
the fields $\phi_1$, $\phi_2$ and $A$, evaluated at zero external
momentum, as well as from tadpole contributions of $\phi_1$ and $\phi_2$. 
Here $\phi_1$, $\phi_2$ and $A$ denote the $CP$-even and $CP$-odd neutral
components of the two Higgs doublets. 

The one-loop corrections to the Higgs pole mass are known without any
approximations~\cite{Chankowski:1991md,Brignole:1992uf, Dabelstein:1994hb}.
The bulk of the numerical effects can be obtained in the so-called
effective-potential approach, for which the external momentum of the Higgs
propagator is set to zero.  Most of the relevant two-loop corrections have
been evaluated in this approach (for reviews, see
e.g. Refs.\,\cite{Heinemeyer:2004ms,Allanach:2004rh}).  Exact calculations at
two-loop order~\cite{Borowka:2014wla,Degrassi:2014pfa} showed explicitely,
that momentum-dependent contributions can reach the current experimental
accuracy only for very heavy supersymmetric particles with masses above few
TeV. In addition, two-loop corrections including even CP-violating couplings
and improvements from renormalization group considerations have been computed
in Refs.~\cite{Heinemeyer:2004ms,Allanach:2004rh,Frank:2006yh}. In particular
CP-violating phases can lead to a shift of a few GeV in $M_h$, see, e.g.,
Refs.~\cite{Heinemeyer:2007aq,Carena:2000yi}.  In Ref.~\cite{Martin:2002wn} a
large class of sub-dominant two-loop corrections to the lightest Higgs boson
mass have been considered.

The first complete three-loop calculation of the leading quartic top
quark mass terms within SUSY QCD has been performed in
Refs.~\cite{Harlander:2008ju,Kant:2010tf} (see also
Ref.~\cite{Mihaila:2013wma} for a recent review). Given the many
different mass parameters entering the formula for the Higgs boson
mass in Eq.~(\ref{eq::MH}) an exact calculation at the three-loop
level is currently not feasible. However, it is possible to apply
expansion techniques~\cite{Smirnov:2002pj} for various limits which
allow to cover a large part of the supersymmetric parameter space.
For the case of the Higgs mass corrections, the occurring Feynman
integrals can be reduced to three-loop tadpole topologies that can be
handled with the program MATAD~\cite{Steinhauser:2000ry}. Concerning
renormalization, it is well known that the perturbative series can
exhibit a bad convergence behaviour in case it is parametrized in
terms of the on-shell top quark mass.  This feature is due to
intrinsically large contributions related to the infra-red behaviour
of the theory. Thus, in Ref.~\cite{Kant:2010tf} the results for the
Higgs boson mass are expressed in terms of the top quark mass
renormalized in the \drbar{} scheme. In addition, in order to avoid
unnatural large radiative corrections for scenarios with heavy
gluinos, a modified non-minimal renormalization scheme for the top
squark masses has also been introduced in Ref.~\cite{Kant:2010tf}.  The
additional finite shifts of top squark masses are chosen such that
they cancel the power-like behaviour of the gluino contributions.

For heavy supersymmetric particles (with a typical mass scale $M_{\rm SUSY}$),
the radiative corrections to the Higgs boson mass contain large logarithms of
the form $\ln(m_t/M_{\rm SUSY})$. They have to be resummed in order to extend
the validity range of the perturbative expansion up to large $M_{\rm SUSY}$
values.  The dominant contributions to the leading (LL) and next-to-leading
logarithmic (NLL) terms up to the fourth loop order have been obtained in
Ref.~\cite{Martin:2007pg}. Very recently, the generalization of the LL and NLL
approximation to the seventh loop order has been derived~\cite{Hahn:2013ria}.
Furthermore, in Ref.~\cite{Draper:2013oza} the recent calculations of the
three-loop beta-functions for the SM couplings\footnote{For details see
  Section~\ref{sec::beta}.}  and the two-loop corrections to the Higgs boson
mass in the SM~\cite{Degrassi:2012ry} have been used to derive the
  (presumably) dominant next-to-next-to-leading logarithmic (NNLL)
corrections at the four-loop order.  Finally, the resummation of the large
logarithms contained in the running of the top quark mass at
three loops within SUSY QCD has been performed in
Ref.~\cite{Kunz:2014gya} and implemented in the code {\tt H3m}.

Usually, the resummation is achieved with the help of Renormalization Group
Equations (RGEs). For example, the resummation performed in
Ref.~\cite{Kunz:2014gya} amounts
to the following steps: the running top quark mass is determined in the SM
with the highest available
precision~\cite{Chetyrkin:1999ys,Chetyrkin:1999qi,Melnikov:2000qh} from the
pole mass. Then, the running mass is evolved up to the
scale where supersymmetric particles become active (SUSY scale) using the RGEs
of the SM. Afterwards, the running top quark mass in the SM is converted to
its value in the MSSM. In this step, threshold corrections at the SUSY scale
are required. In the last step, the running top quark mass in the MSSM is
evolved to the desired energy scale with the help of MSSM RGEs.  As the RGEs
in the SM and the MSSM are known to three-loop order, the threshold
corrections are required at the two-loop order.  The explicit calculations of
the three-loop RGEs and the two-loop supersymmetric threshold corrections are
presented in detail in Section~\ref{sec::beta}.

There are by now several computer programs publicly available which include
most of of the higher order corrections to the lightest Higgs boson mass in
the MSSM.  {\tt FeynHiggs} has been available already since
1998~\cite{FeynHiggs,Frank:2006yh} and has been continuously improved since then. In
particular, its last version contains all numerically important two-loop
corrections, as well as the resummation of the LL and NLL computed in
Ref.~\cite{Hahn:2013ria}.  A second program, {\tt
  CPSuperH}~\cite{Lee:2003nta}, is based on renormalization group improved
calculations and allows for explicit CP violation.  The third program, {\tt
  H3m}~\cite{Kant:2010tf}, contains all currently available three-loop results
together with the resummation of the large logarithms derived
in~\cite{Kunz:2014gya}. In the latest version of {\tt H3m} also the {\tt
  Mathematica} package {\tt SLAM}~\cite{Marquard:2013ita} has been
implemented. {\tt SLAM} provides an interface for calling and reading output
from SUSY spectrum generators fully automatically, using input parameters
specified in the Supersymmetry Les Houches Accord
(SLHA)~\cite{Skands:2003cj}. Furthermore, the Higgs boson masses are also
calculated by the SUSY spectrum generators {\tt
  SoftSusy}~\cite{Allanach:2001kg}, {\tt SPheno}~\cite{Porod:2003um}, and {\tt
  SuSpect}~\cite{Djouadi:2002ze} using $\overline{\rm DR}$ parameters and
two-loop RGEs. For low supersymmetric mass scales below $1$~TeV, the
predictions for $M_h$ of all codes are in quite good agreement with
differences within about $1$~GeV.  This level of agreement is in general
sufficient due to a sizeable parametric uncertainty for the prediction of
$M_h$, that is mainly induced by the experimental uncertainty in the
measurement of the top quark mass. However, for large SUSY scales in the
multi-TeV range, the differences between the predictions of the various codes
becomes substantial. This behaviour can be explained by the increase of the
radiative corrections with the SUSY scale, and by the fact that different
orders in perturbation theory are implemented in the various programs.
 
\begin{figure*}[t]
  \begin{center}
      \includegraphics[angle=0,width=\linewidth]{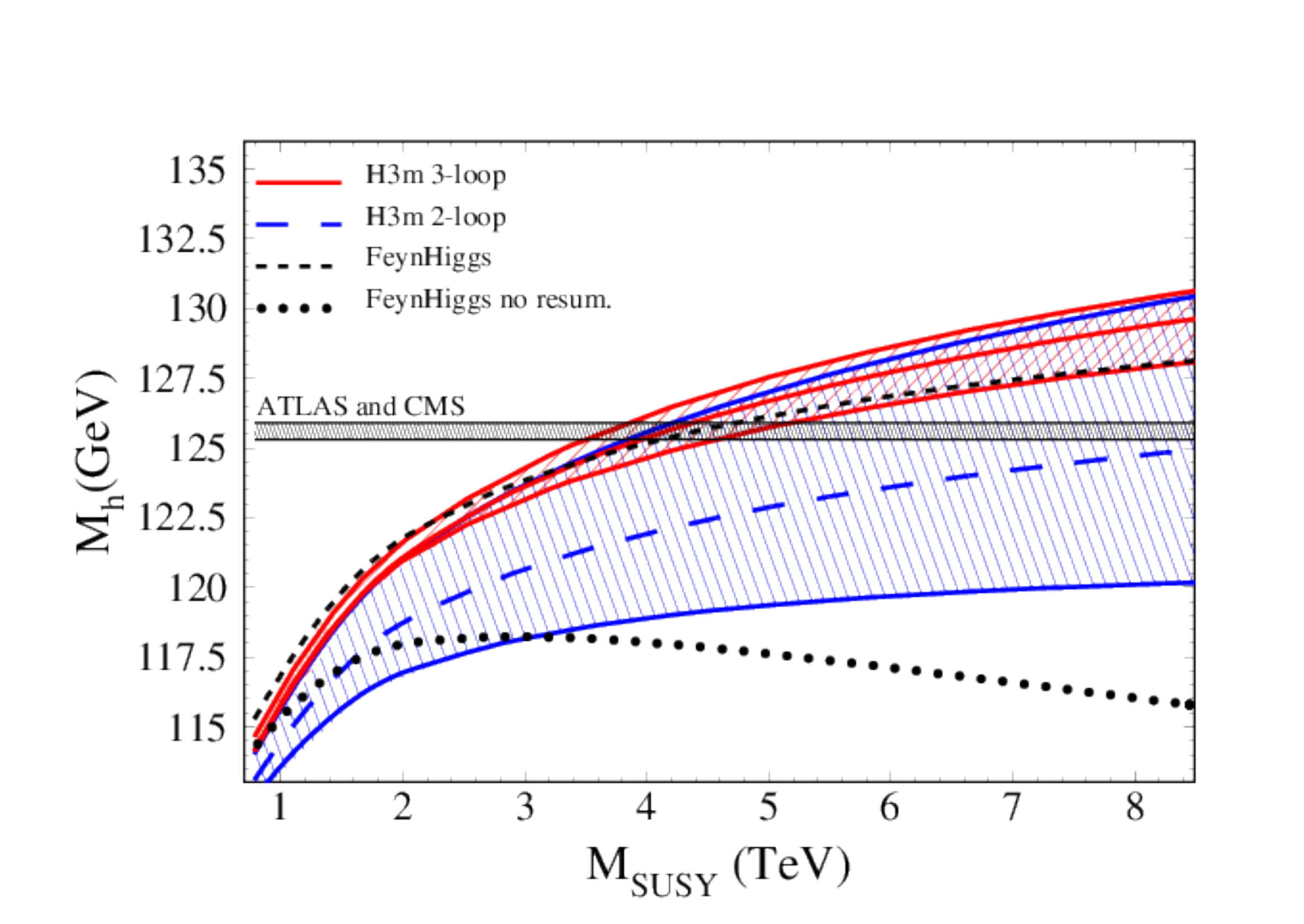}
      \caption[]{\label{fig:mh} Comparison of the predictions for the lightest
        Higgs boson mass provided by {\tt H3m} and {\tt
          FeynHiggs}. The long-dashed and solid lines depict the predictions
        of {\tt H3m} at two- and three-loop orders. The dashed line shows the
        two-loop result obtained with {\tt FeynHiggs} including
        resummation of LL and NLL. The dotted line corresponds to the {\tt
          FeynHiggs} result without resummation.  The error bands account for
        the theoretical uncertainties of SUSY-QCD results induced by the
        variation of the renormalization scale from $m_t/2$ to $M_{\rm
          SUSY}$ (see text for more details).}
  \end{center}
\end{figure*}

  For illustration we show in Fig.~\ref{fig:mh} a comparison of the
  predictions for the Higgs boson mass computed with {\tt H3m} at two and
  three loops (long-dashed and solid lines), and {\tt FeynHiggs} at two
  loops with and without resummation of LL and NLL (dashed and dotted lines).
  The experimentally measured value for the Higgs boson mass is depicted by
  the horizontal black band. 

  For the supersymmetric input parameters we choose
  a constrained MSSM scenario with the following values: $A_0=0$, $\tan\beta =
  10$, $\mu>0$ and $m_0=m_{1/2}=200~\mbox{GeV},\ldots, 5~\mbox{TeV}$ where the
  definition $M_{\rm SUSY}=\sqrt{M_{\rm st_1} M_{\rm st_2}}$ has been used.
  $M_{\rm st_1}$ and $M_{\rm st_2}$ denote the stop quark masses.  For the
  results obtained with {\tt FeynHiggs} and for the central values of the {\tt
    H3m} predictions we fix the renormalization scale to the top quark pole
  mass.  
  The discrepancy between the two-loop
  predictions of the two codes can be traced back to  differences in
  the renormalization schemes and  resummation procedures.
  For the MSSM scenario presented here the {\tt
    FeynHiggs} prediction lies quite close to the central curve of {\tt H3m}
  results at three loops. Note, however, that there are also other choices of
  MSSM parameters where bigger differences between the resummed two-loop
  predictions of {\tt FeynHiggs} and {\tt H3m} are observed.

  The uncertainty bands around the {\tt H3m} predictions have been obtained by
  varying the renormalization scale between $m_t/2$ and $M_{\rm SUSY}$
  using for simplicity the SUSY-QCD corrections up to three-loop order in the
  on-shell scheme as derived in Ref.~\cite{Harlander:2008ju}.  We are aware
  that there is an inconsistency with the central values (long-dashed and
  solid line) which are obtained in the modified \drbar{}
  scheme~\cite{Kant:2010tf}. Nevertheless, the bands should reflect the
  realistic size and variation of uncertainties. Indeed it is nice to see that 
  the three-loop band lies almost completely inside the two-loop band.

  As can be seen from the figure the radiative corrections increase with the
  SUSY mass scale and can amount even at the three-loop level up to few GeV. It
  is important to mention that for heavy SUSY scales, the genuine three-loop
  radiative corrections are few times bigger than the parametric uncertainty
  induced by the uncertainty in the top quark mass measurement (estimated to
  be of around $1$~GeV~\cite{Kant:2010tf}) and about an order of magnitude
  bigger than the current experimental uncertainty on $M_h$. Furthermore,
  guided by the size of theoretical uncertainties at three loops, we conclude
  that to cope with the experimental precision on $M_h$, for heavy SUSY
  particles even the four-loop contributions are required.

  Finally, the determination of the lightest Higgs boson mass including three-loop
  corrections relaxes the lower bound on the SUSY mass scale to about $4$~TeV,
  greatly improving the prospects for supersymmetry discovery at the upcoming
  run of the LHC.  Once again, this numerical analysis highlights the
  importance of improved theoretical calculations of $M_h$ to refine the
  implications of the Higgs boson discovery for constraining the
  supersymmetric models.



\section{\label{sec::ggh}Higgs boson production in the SM and MSSM}

A crucial input for the discovery of a Higgs boson at the LHC was the precise
prediction of the cross section. In fact, there are several production
mechanisms which are nicely summarized in the
reviews~\cite{Dittmaier:2011ti,Dittmaier:2012vm,Heinemeyer:2013tqa}.  In this
section we concentrate on the gluon fusion channel which gives numerically the
largest contribution.

A promising approach to compute higher order corrections to Higgs boson
production is based on the Lagrange density
\begin{eqnarray}
  {\cal L}_{\rm eff} &=& -\frac{H}{v} C_1 \frac{1}{4} G_{\mu\nu} G^{\mu\nu}
  \,,
  \label{eq::leff}
\end{eqnarray}
which describes the effective coupling of a Higgs boson to up to four gluons.
In Eq.~(\ref{eq::leff}) $G_{\mu\nu}$ is the gluon field strength tensor and
$C_1$ is the coupling (or matching coefficient) resulting from integrating out
the heavy degrees of freedom from the underlying theory.  Within QCD, $C_1$
only depends on the top quark mass via $\ln(\mu^2/m_t^2)$ where $\mu$ is the
renormalization scale.  It has been computed up to three loops (NNLO) in
Refs.~\cite{Chetyrkin:1997un,Steinhauser:2002rq,Kramer:1996iq} and the
four-loop calculation has been performed
in~\cite{Schroder:2005hy,Chetyrkin:2005ia}.  Using renormalization group
techniques even the five-loop expression could be derived in
Refs.~\cite{Schroder:2005hy,Chetyrkin:2005ia}, which, however, depends on the
unknown coefficient of the fermionic part of the five-loop beta function.

In the MSSM, $C_1$ becomes a complicated function of
all heavy mass scales and $\mu$, see Subsection~\ref{sub::mssm} for more
details. 

In Refs.~\cite{Harlander:2009mq,Pak:2009dg,Harlander:2009my,Pak:2011hs} it has
been demonstrated that the NNLO prediction of $\sigma(pp\to H+X)$ within the
effective-theory approach of Eq.~(\ref{eq::leff}) approximates the exact SM
result with an accuracy below 1\%, in particular for Higgs boson masses around
$125$~GeV. This issue is discussed in Subsection~\ref{sub::mtcorr}.  Numerical
NLO calculations~\cite{Anastasiou:2008rm} suggest a similar behaviour in the
MSSM.

Although the Higgs production cross section is known to NNLO, the contribution
from unknown higher orders is estimated to be of the order of 10\% which asks
for a N$^3$LO calculation. We summarize the current status in
Subsection~\ref{sub::n3lo}.  Finally, in Subsection~\ref{sub::mssm} we briefly
discuss the production of a light Higgs boson within the MSSM.


\subsection{\label{sub::mtcorr}Top quark mass dependent results up to NNLO}

LO contributions to $\sigma(pp\to H+X)$ have already been computed end of the
seventies in
Refs.~\cite{Wilczek:1977zn,Ellis:1979jy,Georgi:1977gs,Rizzo:1979mf} and also
the NLO QCD corrections are available since almost 20
years~\cite{Dawson:1990zj,Spira:1995rr} taking into account the exact
dependence on the top quark mass (see also Ref.~\cite{Harlander:2005rq} for
analytic results of the virtual corrections). NLO electroweak corrections
have been computed in Ref.~\cite{Actis:2008ug} and mixed QCD-electroweak
corrections are considered in~\cite{Anastasiou:2008tj}.

At LHC energies the NLO QCD corrections amount to 80-100\% of the LO
contributions which makes it mandatory to compute higher order perturbative
corrections.  Beginning of the century three groups have independently
evaluated the NNLO
corrections~\cite{Harlander:2000mg,Harlander:2002wh,Anastasiou:2002yz,Ravindran:2003um}
in the limit of infinitely heavy top quark using the effective Lagrange
density of Eq.~(\ref{eq::leff}). NNLO corrections to the production of a
  pseudo-scalar Higgs boson have been computed in
  Refs.~\cite{Harlander:2002vv,Anastasiou:2002wq}.  At NLO this approximation
works surprisingly well, leading to deviations from the exact result that are
less than 2\% for $M_h<2M_t$ (see, e.g., Ref.~\cite{Harlander:2003xy}).
However, one has to keep in mind that next to $M_h$ and $M_t$ the partonic
cross section also depends on the partonic center-of-mass energy $\hat{s}$
which, in principle, reaches up to the beam energy of the LHC questioning the
validity of the assumption $M_t\to\infty$.  Thus, it is necessary to
investigate the validity of the large top quark mass approximation at NNLO.

A further reason why the effective theory approach needs to be checked at NNLO
is the fact that several improvements over the fixed-order calculation have
been constructed. Among them is the soft-gluon resummation to
next-to-next-to-leading~\cite{Catani:2003zt,Ahrens:2010rs} (see also
Ref.~\cite{Bonvini:2014qga}) and
next-to-next-to-next-to-leading~\cite{Moch:2005ky,Ravindran:2006cg,Catani:2014uta,deFlorian:2014vta}
logarithmic orders and the identification (and resummation) of certain $\pi^2$
terms~\cite{Ahrens:2008nc} which significantly improves the perturbative
series.

\begin{figure}[t]
  \centering
  \includegraphics[width=0.3\linewidth]{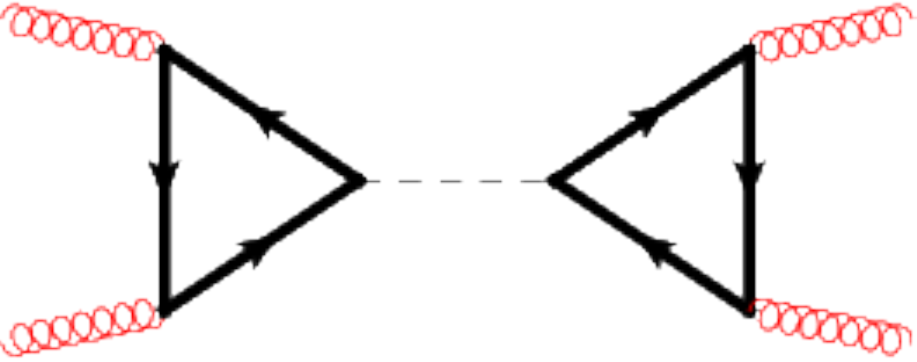}\hfill
  \includegraphics[width=0.3\linewidth]{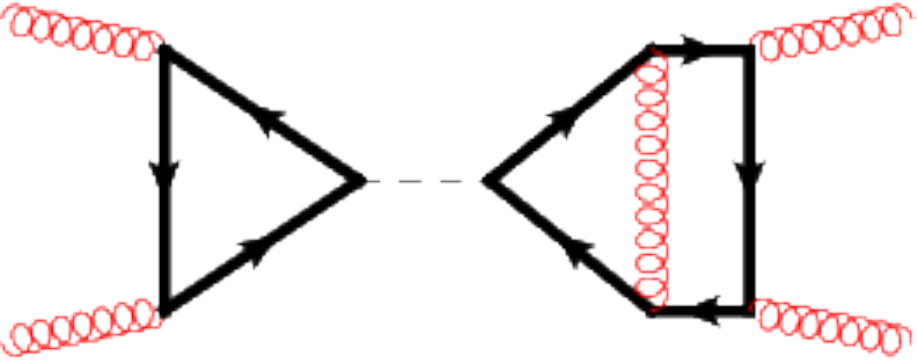}\hfill
  \includegraphics[width=0.3\linewidth]{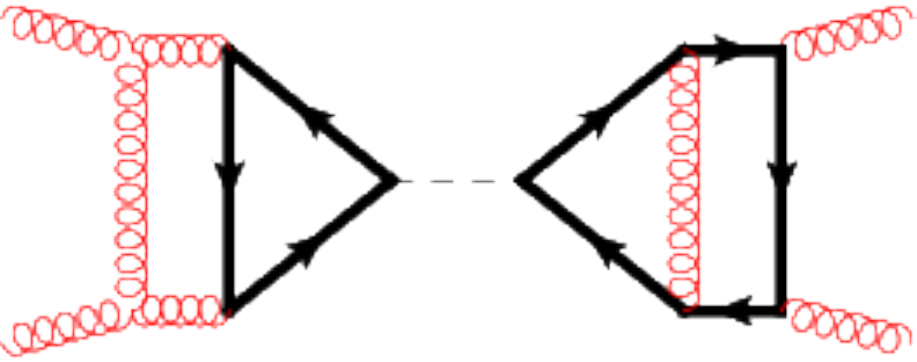}
  \\[1em]
  \includegraphics[width=0.3\linewidth]{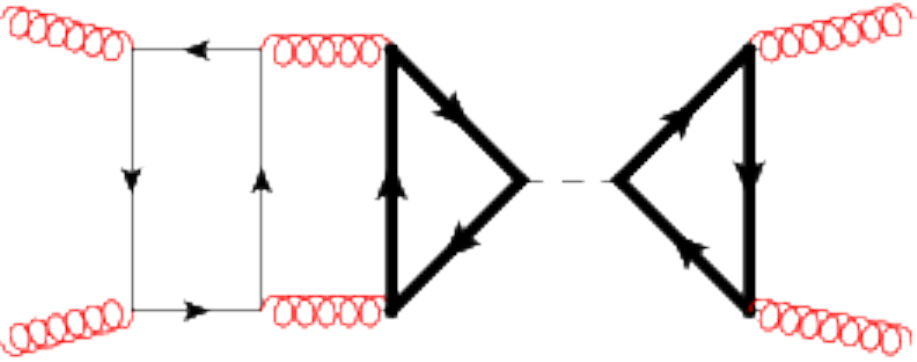}\hfill
  \includegraphics[width=0.3\linewidth]{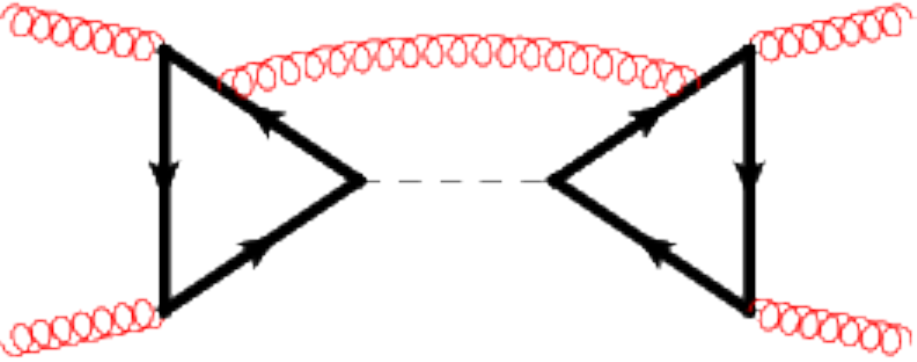}\hfill
  \includegraphics[width=0.3\linewidth]{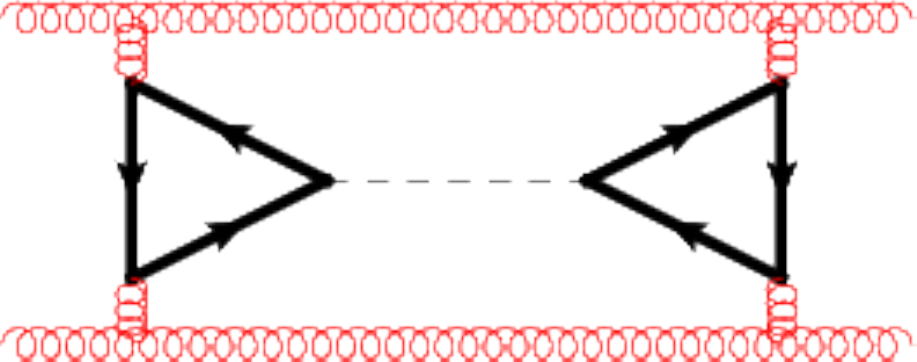}
  \\[1em]
  \includegraphics[width=0.3\linewidth]{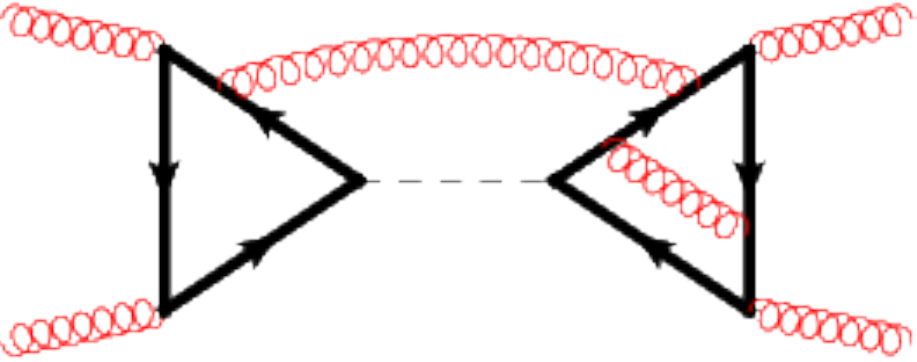}\hfill
  \includegraphics[width=0.3\linewidth]{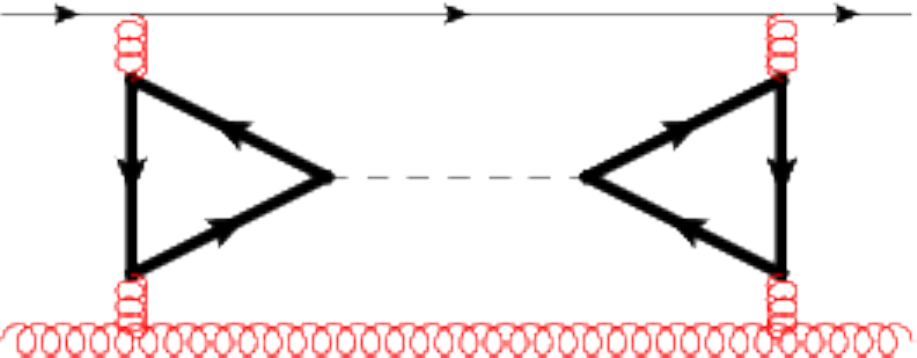}\hfill
  \includegraphics[width=0.3\linewidth]{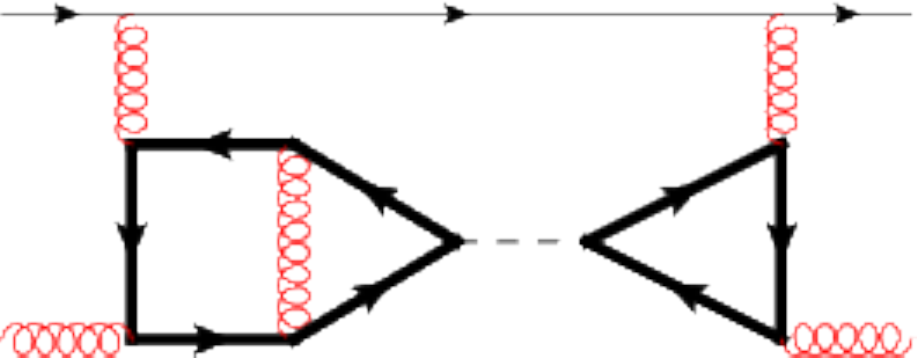}
  \\[1em]
  \includegraphics[width=0.3\linewidth]{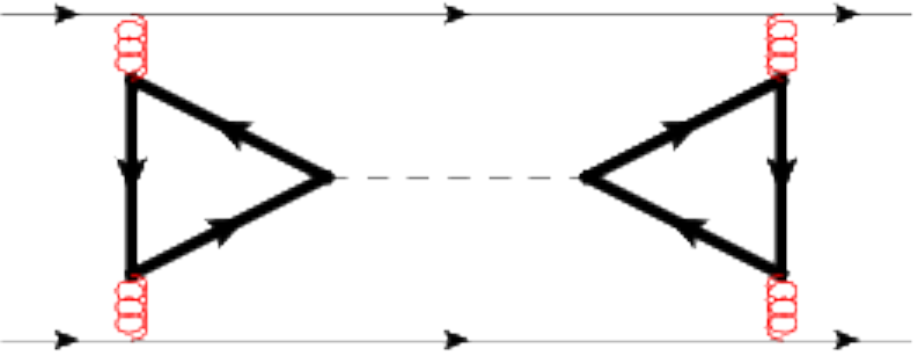}\hfill
  \includegraphics[width=0.3\linewidth]{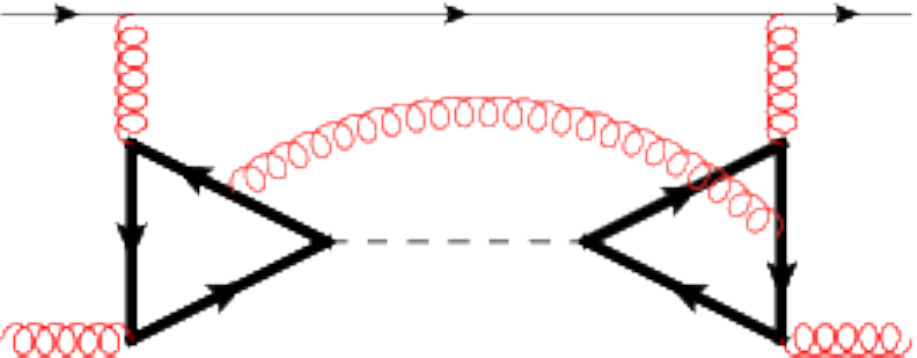}
  \caption[]{\label{fig::gg_n2lo}
    Sample forward scattering diagrams
    whose cuts correspond to the LO, NLO and NNLO corrections to
    $gg\to H +X$, $qg\to H +X$ and $qq\to H + X$. 
    Dashed, curly and thick (thin) solid lines represent
    Higgs bosons, gluons and top (light) quarks, respectively.}
\end{figure}

At NNLO an exact calculation, as it has been performed at NLO, is currently
out of range and thus approximation methods have to be used in order to
estimate the effect of a finite top quark mass. In the analyses performed in
Refs.~\cite{Marzani:2008az,Harlander:2009bw,Pak:2009bx,Harlander:2009mq,Pak:2009dg,Harlander:2009my,Pak:2011hs}
there are actually two ingredients which allow the systematic reconstruction
of an approximation for the partonic cross section:
\begin{itemize}
\item It is suggestive to evaluate the Feynman diagrams contributing to Higgs
  production in the full theory applying an asymptotic expansion for large top
  quark mass. In the approach of Refs.~\cite{Pak:2009dg,Pak:2011hs}, where the
  imaginary parts of forward-scattering amplitudes have been considered, this
  requires the computation of four-loop Feynman diagrams as shown in
  Fig.~\ref{fig::gg_n2lo} for the gluon- and quark-induced channels.  In
  practice, one- or two-loop vacuum integrals with mass scale $M_t$ and one- or
  two-loop box diagrams with massless external particles and a massive Higgs
  boson in forward-scattering kinematics have to be computed.  As a result one
  obtains the partonic cross section in an expansion in the inverse top quark
  mass which should provide a good approximation for $M_t^2\gg M_h^2,\hat{s}$.
  Four and five expansion terms in $1/M_t^2$ have been computed for a scalar
  and pseudo-scalar Higgs boson, respectively.
\item 
  The second ingredient is the leading high-energy behaviour for the partonic
  cross section obtained in Refs.~\cite{Marzani:2008az,Harlander:2009my}
  and~\cite{Caola:2011wq} for the scalar and pseudo-scalar Higgs boson,
  respectively.
  At NLO it is given by a constant, at NNLO, however, the leading
  term is proportional to $\ln(x)$ (with $x = M_h^2/\hat{s}$)
  and the constant is not known.
\end{itemize}

\begin{figure}[t]
  \centering
  \includegraphics[width=0.9\linewidth]{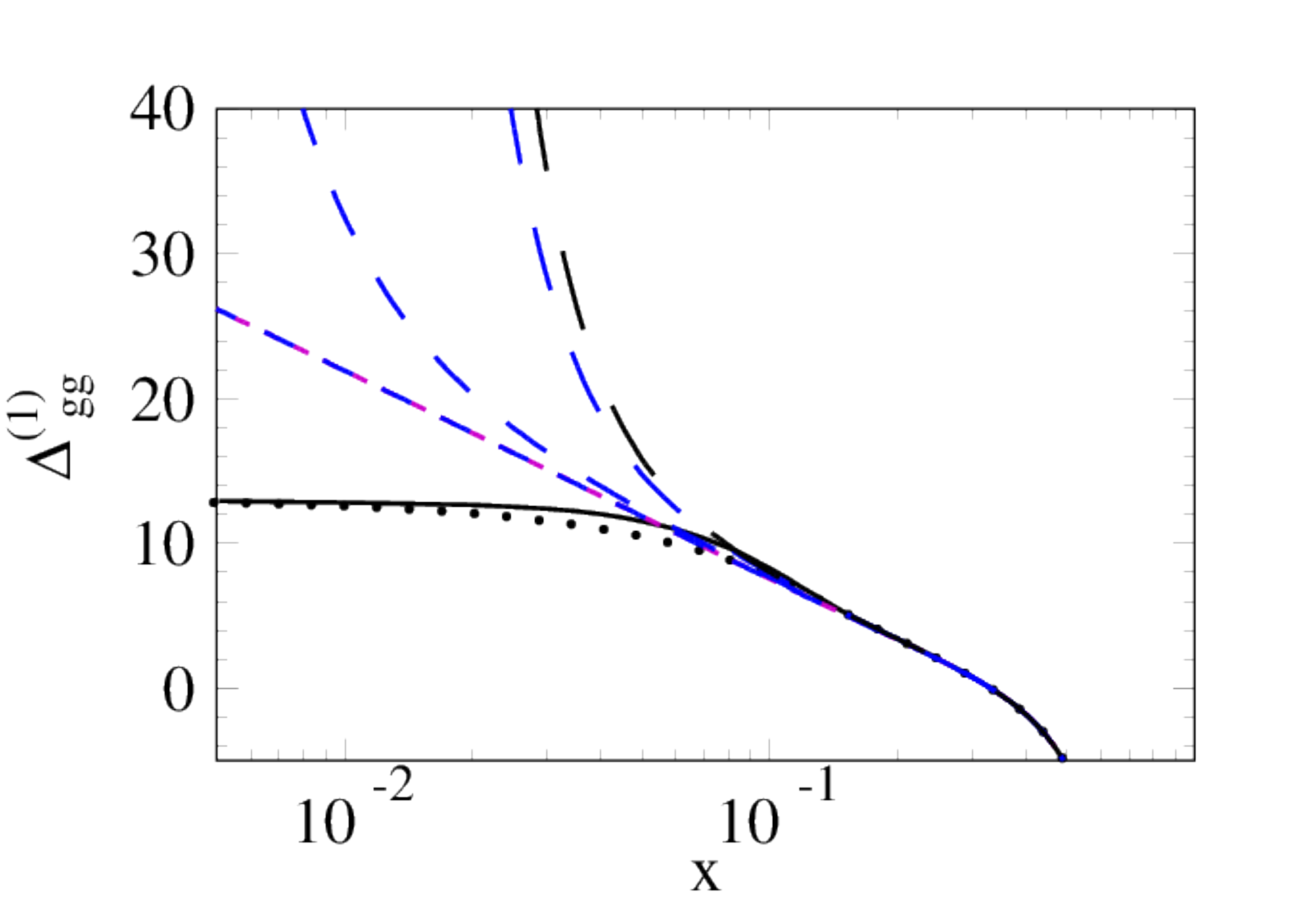}
  \caption[]{\label{fig::gg_nlo}
    Partonic NLO cross sections for the gg  channel as
    function of $x=M_h^2/\hat{s}$ for $M_h = 130$~GeV.
    The exact and approximated results are shown as solid and dotted lines and
    the dashed lines correspond to expansions in $1/M_t$ (long-dashed curves 
    contain more expansion terms).}
\end{figure}

In Refs.~\cite{Marzani:2008az,Caola:2011wq} the high-energy behaviour has been
matched to the infinite-top quark mass result.  A systematic study taking into
account higher order $1/M_t$ results has been performed in
Refs.~\cite{Harlander:2009mq,Pak:2009dg,Harlander:2009my,Pak:2011hs}.  To
illustrate the procedure we show in Fig.~\ref{fig::gg_nlo} the partonic cross
section at NLO as a function of $x = M_h^2/\hat{s}$ for
$M_h=130$~GeV~\cite{Pak:2009dg}. The quantity
$\Delta_{gg}^{(1)}$ is defined via~\cite{Pak:2009dg}
\begin{eqnarray}
  \hat{\sigma}_{gg\to H + X} &=& \hat{A}_{\rm LO} \left(
    \Delta_{gg}^{(0)} + \frac{\alpha_s}{\pi}~\Delta_{gg}^{(1)}  
    + \ldots
  \right)
  \,,
  \label{eq::hatsigma}
\end{eqnarray}
where $\hat{A}_{\rm LO}$ collects various constants and the exact top quark
mass dependence. Lines with longer dashes include higher
order terms in the $1/M_t$ expansion which only converges up to the threshold
at $x \approx 0.14$. Nevertheless, the approximation constructed in
Ref.~\cite{Pak:2009dg} (dotted line) agrees well with
the exact result (solid line, obtained from {\tt HIGLU}~\cite{Spira:1995mt})
and leads to a negligible deviation for the hadronic cross section.  Indeed,
even the difference in the hadronic cross section computed from the
approximated result and the $M_t\to\infty$ expression (short-dashed
straight line for $x\to0$ in Fig.~\ref{fig::gg_nlo}.) is below $1\%$ at 
NLO for a scalar Higgs boson with $M_h=125$~GeV and of the order
of 6\% for a pseudo-scalar Higgs boson with mass $M_A=300$~GeV.

A similar behaviour is observed at NNLO: the difference 
between the effective-theory result and the
hadronic cross section computed for finite top quark mass is below 1\%
for $M_h=125$~GeV and can amount to about 10\% for a pseudo-scalar
Higgs boson with mass $M_A=300$~GeV. Thus, for a scalar Higgs boson
as observed at the LHC finite top quark mass effects are negligible beyond Born
approximation for the inclusive cross section.


\subsection{\label{sub::n3lo}Status at NNNLO and beyond}

\begin{figure}[t]
  \begin{center}
    \begin{tabular}{ccc}
      \includegraphics[width=.28\linewidth]{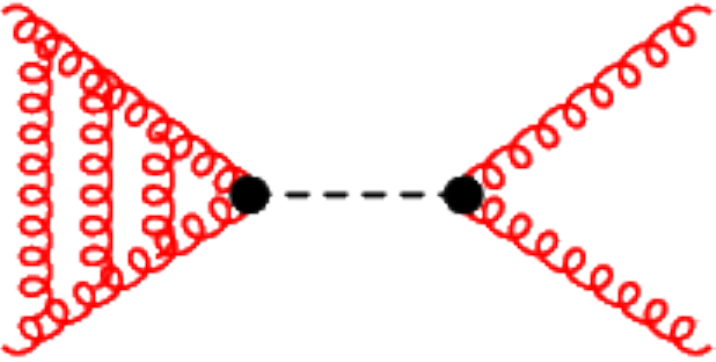} &
      \includegraphics[width=.28\linewidth]{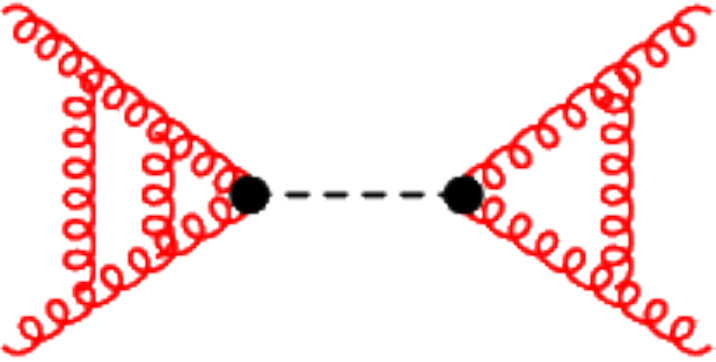} &
      \includegraphics[width=.28\linewidth]{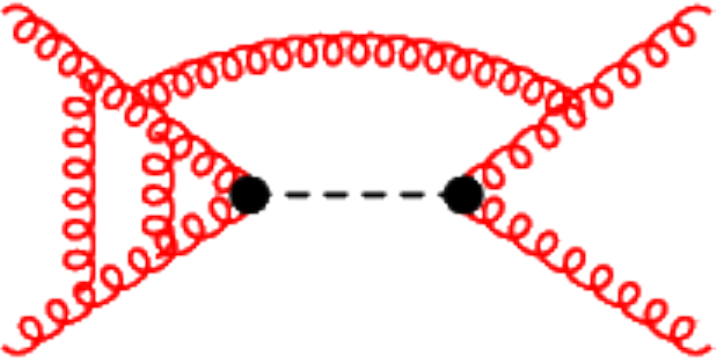} \\
      (a) & (b) & (c) \\[1em]
      \includegraphics[width=.28\linewidth]{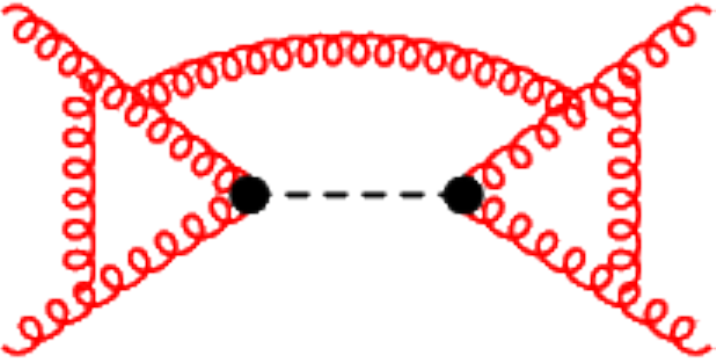} &
      \includegraphics[width=.28\linewidth]{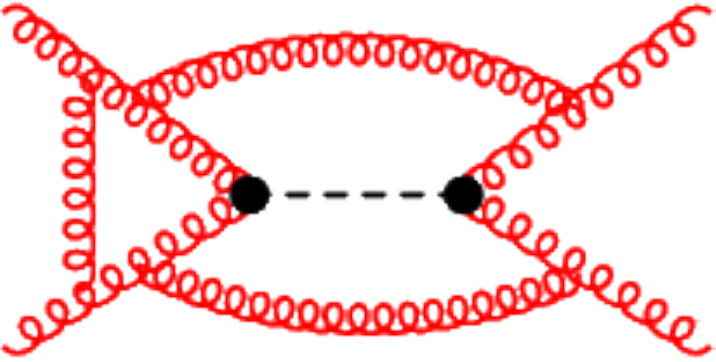} &
      \includegraphics[width=.28\linewidth]{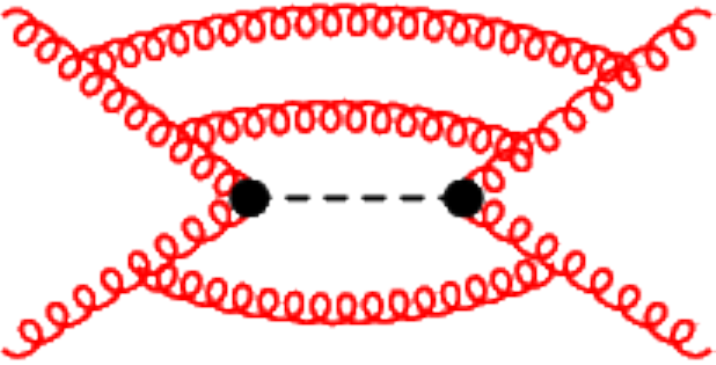} \\
      (d) & (e) & (f)
    \end{tabular}
    \caption{\label{fig::n3lo}Sample Feynman diagrams in forward scattering
      kinematics contributing to the N$^3$LO cross section for Higgs boson
      production in gluon fusion.  All cuts through the Higgs boson line have
      to be considered.  Dashed and curly lines correspond to Higgs bosons and
      gluons, respectively. The black blob indicates the effective Higgs-gluon
      vertex according to Eq.~(\ref{eq::leff}).}
  \end{center}
\end{figure}

The various contributions which have to be considered for $\sigma(pp\to H+X)$
at N$^3$LO are shown in Fig.~\ref{fig::n3lo} where amplitudes with
forward-scattering kinematics are shown. From these diagrams all cuts through
the Higgs boson lines have to be computed. They include three-loop virtual
corrections [see (a) and (b)], two-loop virtual corrections in association
with a real emission of a parton [see (c)], squared contribution of the
one-loop real-virtual corrections [see (d)], the one-loop virtual contribution
with two real emissions [see (e)], and the contribution with three real
emissions [see (f)]. In addition collinear counterterms for the parton
distribution functions have to be taken into account in order to arrive at an
infrared finite quantity.  Different groups have provided building blocks for
the N$^3$LO Higgs boson production cross section which will be briefly
summarized in the following.  

\begin{itemize}

\item In Ref.~\cite{Chetyrkin:1997un} the four-loop corrections to the
  matching coefficient $C_1$ of the effective Lagrangian~(\ref{eq::leff}) have
  been constructed from the three-loop decoupling constant for the strong
  coupling constant with the help of renormalization group methods and a
  low-energy theorem (see also Section~\ref{sub::LET}). In
  Refs.~\cite{Schroder:2005hy,Chetyrkin:2005ia} the result has been confirmed
  by an explicit calculation of the four-loop decoupling constant.

\item The three-loop corrections to the massless Higgs-gluon
  [cf. Fig.~\ref{fig::n3lo}(a) and (b)] (and photon-quark) form factor have
  been obtained by two independent
  calculations~\cite{Baikov:2009bg,Gehrmann:2010ue} (see also
  Ref.~\cite{Lee:2010cga};
  fermionic contributions to the photon-quark form factor 
  have already been obtained in Ref.~\cite{Moch:2005tm}).

\item The ${\cal O}(\epsilon)$ contributions to the NNLO master integrals 
  has been computed in Refs.~\cite{Pak:2011hs,Anastasiou:2012kq}.

\item Results for the LO, NLO and NNLO partonic cross sections expanded up to
  order $\epsilon^3$, $\epsilon^2$ and $\epsilon^1$, respectively, 
  have been published in Refs.~\cite{Hoschele:2012xc,Buehler:2013fha}.

\item All contributions from convolutions of partonic cross sections with
  splitting functions, which are needed for the complete N$^3$LO calculation,
  are provided in Refs.~\cite{Hoschele:2012xc,Hoeschele:2013gga}.  The results
  of~\cite{Hoschele:2012xc} have been confirmed in
  Ref.~\cite{Buehler:2013fha}.

\item The full scale dependence of the N$^3$LO expression has
  been constructed in Ref.~\cite{Buehler:2013fha}.

\item 
  The single-soft current to two-loop order has been computed in
  Refs.~\cite{Duhr:2013msa,Li:2013lsa} which is an important ingredient to the
  two-loop corrections with one additional real radiation
  [cf. Fig.~\ref{fig::n3lo}(c)]. 
  The latter have been computed in Refs.~\cite{Dulat:2014mda,Duhr:2014nda}.

\item
  The single-real radiation contribution which originates from the square of
  one-loop amplitudes [see Fig.~\ref{fig::n3lo}(d)] has been computed to all
  orders in $\epsilon$ in Refs.~\cite{Anastasiou:2013mca,Kilgore:2013gba}.

\item
  The soft limit of the 
  phase space integrals for Higgs boson production in association with two
  soft partons [cf. Fig.~\ref{fig::n3lo}(e)] were computed in
  Refs.~\cite{Anastasiou:2014vaa,Li:2014bfa}, in the latter reference even to all
  orders in $\epsilon$.

\item The triple-real contribution to the gluon-induced partonic cross section
  [cf. Fig.~\ref{fig::n3lo}(f)] has been considered in
  Ref.~\cite{Anastasiou:2013srw} in the soft limit, i.e. for $y = 1 -
  M_h^2/\hat{s}\to 0$. In particular, a method has been developed which allows the
  expansion around the soft limit. Two expansion terms in $y$ are provided.

\item
  Three-loop ultraviolet counterterms are needed for
  $\alpha_s$~\cite{Tarasov:1980au,Larin:1993tp} and the effective
  operator~\cite{Spiridonov:1984br}.

\item Two leading terms in the threshold expansion for the complete N$^3$LO
  total Higgs production cross section through gluon fusion has been presented
  in Refs.~\cite{Anastasiou:2014vaa,Anastasiou:2014lda,Li:2014afw}. In these
  references, for the first time, a complete third-order cross section has
  been constructed (although only for $y\to 0$) which constitutes an
  important step. For physical applications probably more terms in the threshold
  expansion are necessary~\cite{Anastasiou:2014lda}.

\item
  A further activity concerns the development of systematic approaches to 
  compute the master integrals for $\sigma(pp\to H+X)$, see, e.g.,
  Refs.~\cite{Anastasiou:2013srw,Hoschele:2014qsa}. 

\item Several groups have constructed approximate N$^3$LO results for the total
  cross section taking into account information from soft-gluon
  approximation and the high-energy
  limit~\cite{Moch:2005ky,Ahrens:2010rs,Ball:2013bra,Bonvini:2014jma,Bonvini:2014joa,Catani:2014uta,deFlorian:2014vta}. Differences
  in the numerical results can partly be traced back to different procedures used
  for the resummation of higher order logarithmic contributions.

\end{itemize}
Several results also apply to third order
corrections to the Drell-Yan process; see, e.g., Ref.~\cite{Ahmed:2014cla}.


\subsection{\label{sub::mssm}Higgs boson production in the MSSM}

A Higgs boson with a mass of about 125~GeV is both consistent with the SM
taking into account the available precision data (see, e.g., Ref.~\cite{Baak:2014ora})
but also with supersymmetric theories, in particular the MSSM (see
Section~\ref{sec:mh}). Thus, it is important to investigate also the quantum
corrections in this theory.  In the recent years several groups have provided
significant contributions in this respect, mainly in the effective-theory
framework which requires the computation of loop corrections to the matching
coefficient $C_1$ and hence only vacuum integrals have to be
considered. Important two-loop contributions have been calculated in
Refs.~\cite{Harlander:2003bb,Harlander:2004tp,Degrassi:2008zj,Anastasiou:2008rm,Degrassi:2010eu,Harlander:2010wr}.

At NLO there are also considerations in the full theory.  In
Ref.~\cite{Anastasiou:2008rm} the total cross section has been computed
numerically considering both the top and bottom sector and building blocks for
a (semi) analytic full-theory calculation have been provided in
Refs.~\cite{Muhlleitner:2006wx,Bonciani:2007ex}; a complete calculation along
these lines is still missing.\footnote{See Ref.~\cite{Muhlleitner:2010nm} for
  preliminary results.}

Recently the effective-theory NLO corrections have been implemented in the
publicly available computer code {\tt SusHi}~\cite{Harlander:2012pb}. At NNLO
the rough approximation of Ref.~\cite{Harlander:2003kf} has been implemented,
i.e., the genuine supersymmetric corrections to $C_1$ have been set to
zero at three loops.  A comprehensive summary of all available contributions
towards a precise prediction of the Higgs boson production cross section is
provided in Ref.~\cite{Bagnaschi:2014zla} which also contains a detailed
discussion of the theoretical uncertainties.

\begin{figure*}
  \centering
  \includegraphics[width=.9\linewidth]{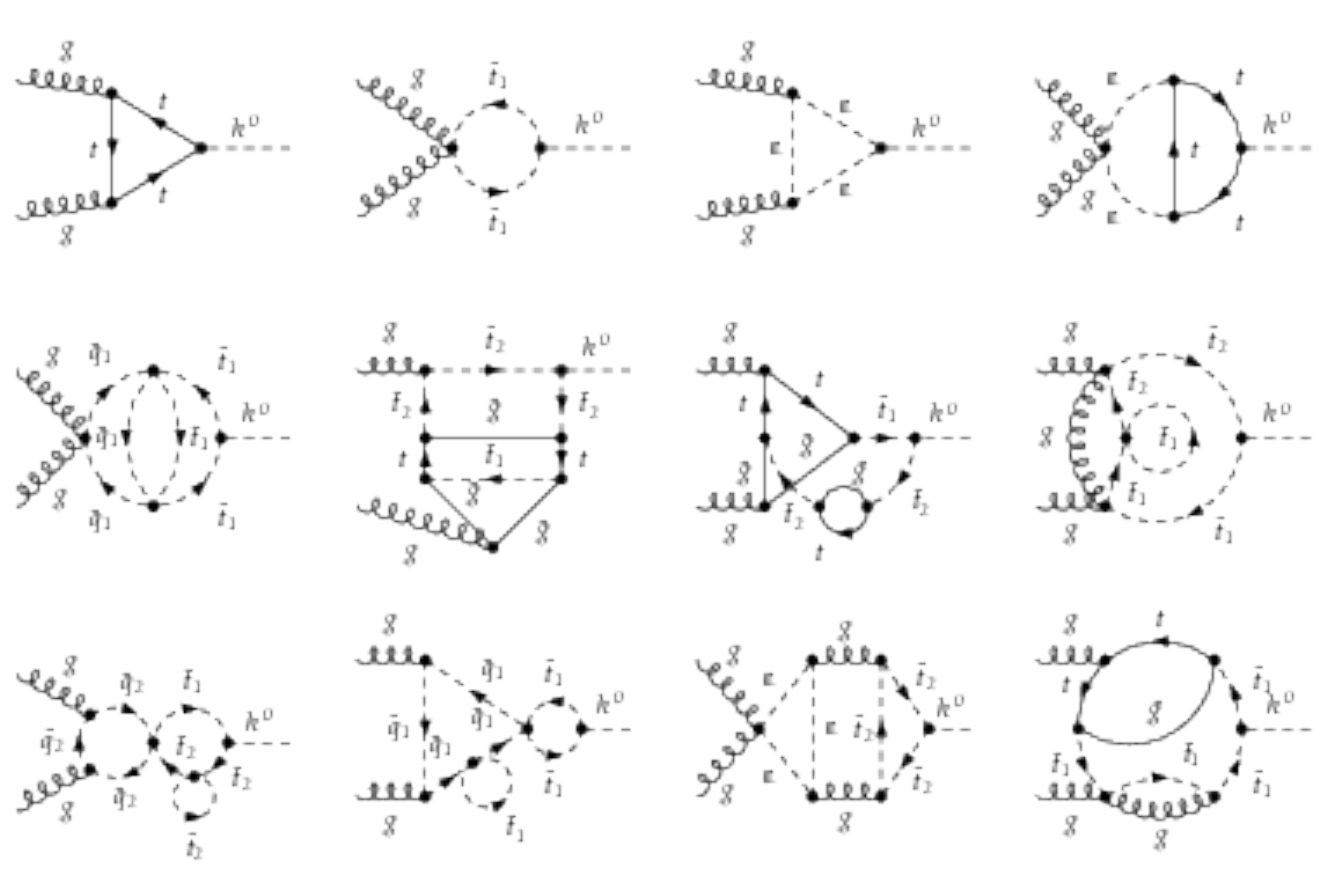}
  \caption{Feynman diagrams contributing to $C_1$ within the MSSM.}
  \label{fig::diags1} 
\end{figure*}

In the remaining part of this section we will discuss the NNLO corrections to
$C_1$ which have been computed within the CRC/TR~9.  Sample diagrams
contributing to $C_1$ at one, two and three loops are shown in
Fig.~\ref{fig::diags1}.  The symbols $t$, $\tilde{t}_i$, $g$, $\tilde{g}$, $h$
and $\varepsilon$ denote top quarks, top squarks, gluons, gluinos, Higgs
bosons and $\varepsilon$ scalars, respectively. The latter are auxiliary
particles introduced to implement regularization by Dimensional
Reduction~\cite{Siegel:1979wq}  which respects supersymmetry.  For the
computation of $C_1$ it is possible to expand the Feynman integrals in the
external gluon momenta which leads to vacuum integrals. In contrast to the SM,
in the MSSM many different mass scales are present which increases
significantly the
complexity of the calculation. In fact, the currently available
tools do not allow for an exact calculation and one has to rely on
approximation methods. In Refs.~\cite{Pak:2012xr,Kurz:2012ff} relations
between the various masses have been assumed such that phenomenologically
interesting scenarios can be studied.  This includes both strong hierarchies
among masses but also expansions in the mass differences.  Note that in the
latter case the expansion series can be written down in different but
equivalent ways. For example, for $x\approx1$ (where $x$ stands for a ratio of
masses) one can choose $1-x$, $1-1/x$, $1-x^2$ or $1-1/x^2$ as expansion
parameter which all lead to the same result after considering the expansion up
to a fixed order. However, in practice different choices lead to different
numerical results and also show different convergence properties (see also
Ref.~\cite{Eiras:2006xm}).  In the calculation of
Refs.~\cite{Pak:2012xr,Kurz:2012ff} up to five different masses occur which
leads (for the hierarchies considered in Ref.~\cite{Pak:2012xr}) to up to 48
different possible representations.  We have implemented sophisticated
expansion schemes with the purpose to select the representation with largest
radius of convergence providing at the same time reliable error estimates for
each point in the parameter space (see Refs.~\cite{Pak:2012xr,Zerf:2012} for a
detailed discussion of the algorithm).  In this way three-loop corrections to
$C_1$ have been evaluated for the top and bottom sector, neglecting, however,
the bottom Yukawa coupling.

In the MSSM the coupling of the light CP-even Higgs boson to bottom quarks is
proportional to $\tan\beta$ whereas the top quark-Higgs coupling is
proportional to $1/\tan\beta$. Thus, for large values of $\tan\beta$ (in
practice this means $\tan\beta\ge10$; see, e.g., Ref.~\cite{Harlander:2010wr})
the bottom sector can\footnote{Whether it indeed leads to large
  corrections depends on the values of the other parameters like the mixing
  angles in the Higgs sector.} contribute significantly to Higgs boson
production. The computation of the Feynman integrals with internal bottom
quark loops are more challenging since it is not possible to apply
an effective theory approach as for the top sector. Indeed, up to now only NLO
corrections are available, both for the SM~\cite{Spira:1995rr}
and for SUSY QCD~\cite{Anastasiou:2008rm,Degrassi:2010eu,Harlander:2010wr}.

At this point we want to mention some
field-theoretical issues connected to the
$\varepsilon$ scalar which have been addressed
in the course of the calculation
performed in Ref.~\cite{Pak:2012xr}. In fact, besides a mass term for the
$\varepsilon$ scalar also a coupling of two $\varepsilon$ scalars to a Higgs
boson, which emerges through radiative corrections, has to be added to the
Lagrange density. Its non-standard part then reads
\begin{eqnarray}
  {\cal L}_\varepsilon &=& 
  -\frac{1}{2}\left(M_\varepsilon^0\right)^2 
  \varepsilon^{0,a}_\sigma\varepsilon^{0,a}_\sigma
  - \frac{\phi^0}{v^0} \left(\Lambda_\varepsilon^0\right)^2
  \varepsilon^{0,a}_\sigma\varepsilon^{0,a}_\sigma 
  \,,
  \label{eq::Lep}
\end{eqnarray}
where $\varepsilon^{0,a}$ denotes the bare $\varepsilon$ scalar field and the
dimensionful quantity $\Lambda_\varepsilon^0$ mediates the coupling of the
Higgs boson to $\varepsilon$ scalars.  It is convenient to renormalize the
$\varepsilon$ scalar mass on-shell requiring $M_\varepsilon^{\rm OS}=0$,
and $\Lambda_\varepsilon^0$ via the
following condition
\begin{eqnarray}
  \left(\Lambda_\varepsilon^0\right)^2
  &=&
  \delta\Lambda_{\varepsilon}^2 + \Lambda_\varepsilon^2
  \,,
\end{eqnarray}
where $\delta\Lambda_{\varepsilon}^2$ is fixed via the condition that
the renormalized coupling of the $\varepsilon$ scalars to Higgs bosons is zero.
Analytic results for the corresponding counterterms, which are needed up to
the two-loop level in the case of $\Lambda_\varepsilon$, can be found in
Ref.~\cite{Pak:2012xr}.

In a first step a simplified scenario with degenerate supersymmetric
masses has been considered in~\cite{Pak:2010cu}. In Ref.~\cite{Pak:2012xr} the
results have been generalized by considering various hierarchies of the
involved supersymmetric particle masses.  Furthermore, details on the
renormalization procedure and the treatment of evanescent couplings have been
discussed. The results of~\cite{Pak:2012xr} have been cross-checked in
Ref.~\cite{Kurz:2012ff} where a low-energy theorem has been used in order to
obtain $C_1$ from the decoupling constant of $\alpha_s$.

\begin{figure*}[t]
  \centering
  \mbox{\includegraphics[width=.47\linewidth]{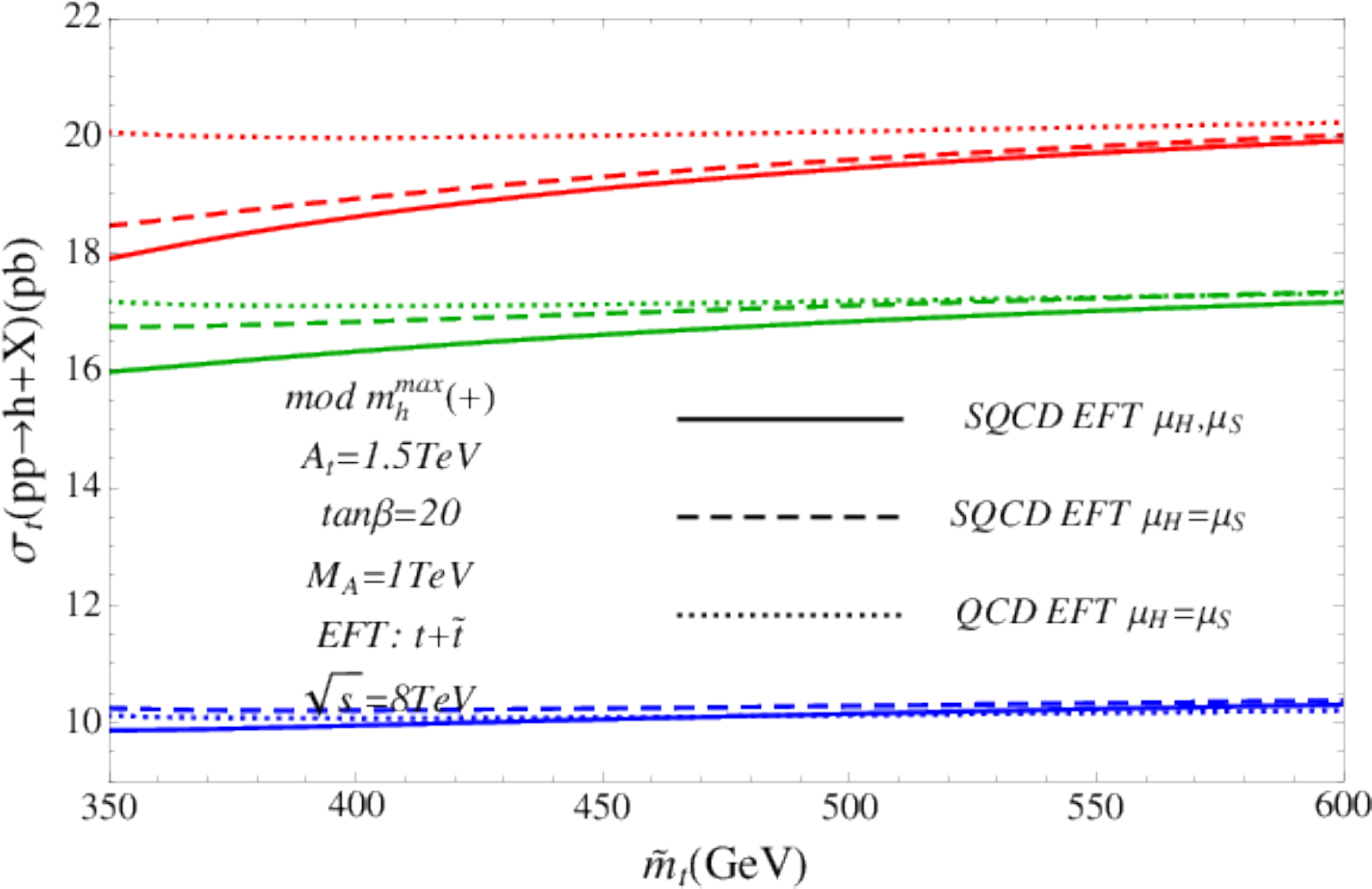}
    \qquad  \includegraphics[width=.47\linewidth]{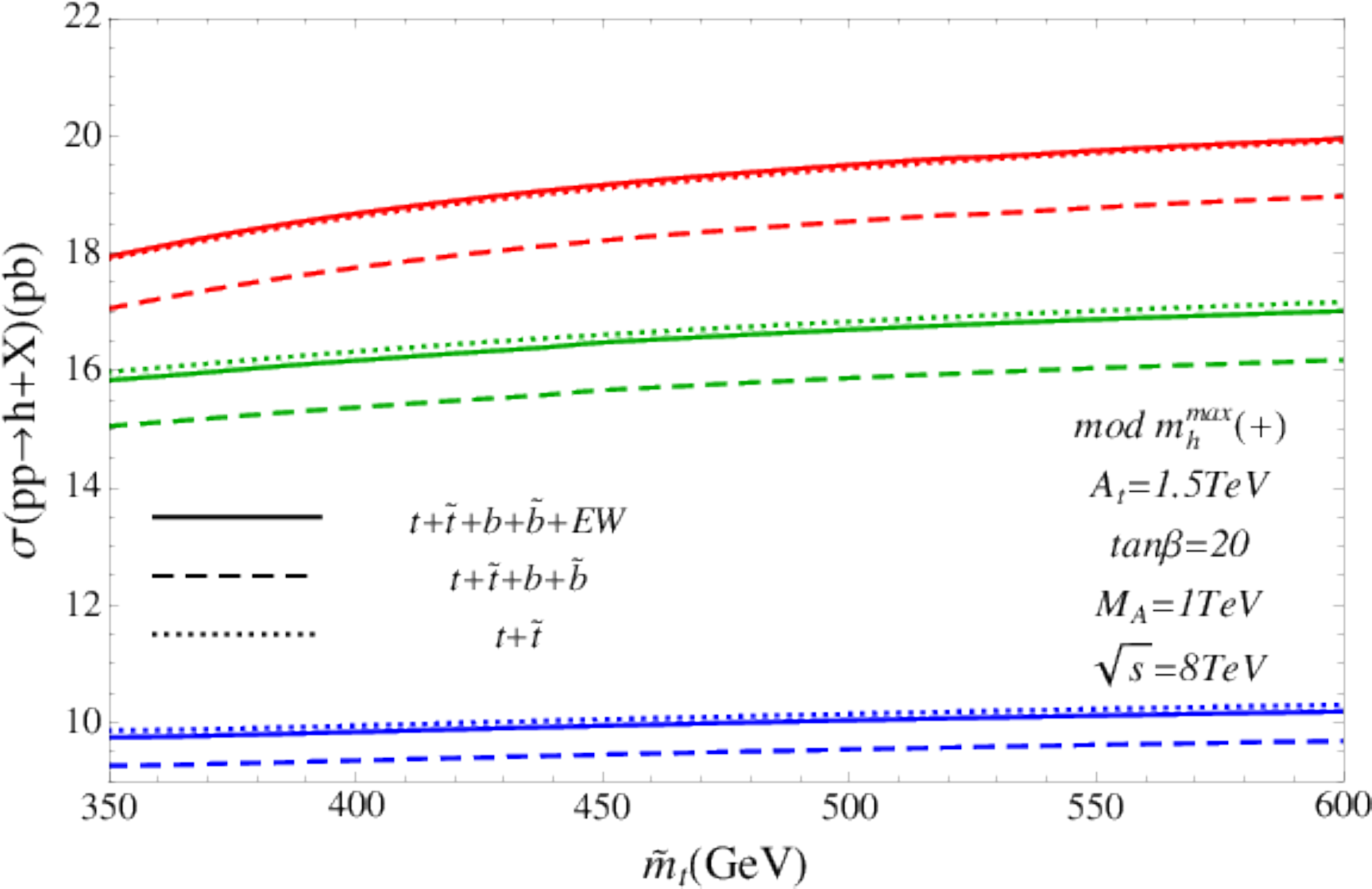}
  }
  \\ (a) \hspace*{22em} (b)
  \caption{Cross section as a function of the singlet soft SUSY breaking
    parameter of the right-handed top squark, $\tilde{m}_t$. (a) top quark/top
    squark contribution $\sigma_t$. (b) complete contribution including also
    bottom quark and electroweak effects as described below
    Eq.~(\ref{eq::sigma}). Figure taken from Ref.~\cite{Pak:2012xr}.}
  \label{fig::sig_mssm}
\end{figure*}

In Fig.~\ref{fig::sig_mssm} we present results for the total cross section 
which is computed according to the formula
\begin{eqnarray}
  \label{eq::sigma}
  \lefteqn{\sigma(pp\to h+X) \,=\, \left( 1 + \delta^{\rm EW} \right) 
  \times}
  \nonumber  \\&&
  \Bigg[\sigma_{tb}^{\rm SQCD}(\mu_s)\bigg|_{\rm NLO}
    -   \sigma_t^{\rm SQCD}(\mu_s)\bigg|_{\rm NLO}
  \nonumber  \\&&
    +   \sigma_t^{\rm SQCD}(\mu_s,\mu_h)\bigg|_{\rm NNLO}
  \Bigg]
  \,,
\end{eqnarray}
where $\sigma_{tb}^{\rm SQCD}$ refers to the NLO cross section including all
top and bottom effects~\cite{Degrassi:2010eu,Harlander:2010wr}. After subtracting
the top quark/top squark contributions with the help of $\sigma_t^{\rm
  SQCD}(\mu_s)|_{\rm NLO}$ we can add the result from the top quark/top squark
up to NNLO.  Note that $\sigma_t$ also contains numerically small
contributions from a non-vanishing Higgs-bottom squark coupling whereas the
bottom Yukawa coupling is set to zero, see Ref.~\cite{Pak:2012xr} for details.
Finally, electroweak effects~\cite{Actis:2008ug} are taken into account in
a multiplicative way.  Note that they are only available in the SM
and potential large MSSM effects are neglected in Eq.~(\ref{eq::sigma}).

In Fig.~\ref{fig::sig_mssm} we discuss numerical effects of the individual
terms in Eq.~(\ref{eq::sigma}) using the $m_h^{\rm max}$ scenario of
Ref.~\cite{Carena:2002qg} as a basis. We apply slight modifications which
lead to the following parameters (see Ref.~\cite{Pak:2012xr} for explanations
of the parameters)
\newcommand{\muSUSY}{\mu_{\rm susy}}
\newcommand{\softsusy}{SOFTSUSY}
\newcommand{\mgluino}{m_{\tilde{g}}}
\newcommand{\mstopone}{m_{\tilde{t}_1}}
\newcommand{\mstoptwo}{m_{\tilde{t}_2}}
\newcommand{\msquark}{m_{\tilde{q}}}
\begin{align}
  &&A_b=A_\tau=2469.48~\mbox{GeV}\,,
  &&A_t=1500~\mbox{GeV}\,,
  \nonumber\\
  &&M_1 = 5 s_W^2 / (3 c_W^2) M_2\,,
  &&M_2 = 200~\mbox{GeV}\,,
  \nonumber\\
  &&M_3 = 800~\mbox{GeV}\,,
  &&M_A=1000~\mbox{GeV}\,,
  \nonumber\\
  &&\muSUSY=200~\mbox{GeV}\,,
  &&m_{\rm susy}=1000~\mbox{GeV}\,,
  \nonumber\\
  &&\tan\beta = 20\,.
  \label{eq::parameters3}
\end{align}
In addition we have the parameter
$\tilde{m}_t$, the soft SUSY breaking parameter of the
top squark, which is varied in Fig.~\ref{fig::sig_mssm}.
The default value $\tilde{m}_t=400$~GeV in combination with
{\tt \softsusy}~\cite{Allanach:2001kg} leads to the following values for the
$\overline{\rm DR}$ masses
\begin{align}
  & \mstopone = 370~\mbox{GeV}\,, & \mstoptwo = 1045~\mbox{GeV}\,, & 
  \nonumber\\
  & \msquark = 1042~\mbox{GeV}\,, & \mgluino = 860~\mbox{GeV}\,, & 
\end{align}
where $\msquark$ corresponds to the average of $m_{\tilde{u}}$,
$m_{\tilde{d}}$, $m_{\tilde{s}}$, $m_{\tilde{c}}$ and $m_{\tilde{b}}$ and the
renormalization scale has been set to the on-shell top quark mass.

In Ref.~\cite{Pak:2012xr} the program {\tt
  H3m}~\cite{Harlander:2008ju,Kant:2010tf} has been used in order to compute
the lightest MSSM Higgs boson mass.  Combining {\tt H3m} with version~2.6.5 of
{\tt FeynHiggs}~\cite{Frank:2006yh} and version~3.1.1 of {\tt
  \softsusy}~\cite{Allanach:2001kg} leads to a Higgs boson mass of
approximately 126~GeV almost independent of $\tilde{m}_t$~\cite{Pak:2012xr}.

In Fig.~\ref{fig::sig_mssm}(a) the quantity $\sigma_t^{\rm SQCD}$ is shown as
a function of $\tilde{m}_t$ at LO, NLO and NNLO (from bottom to top). For each
order three curves are shown where the dotted curve corresponds to the
SM.
The SUSY QCD corrections are included in the dashed and solid line
where for the former the soft and hard renormalization scales, $\mu_s$ and
$\mu_h$ have been identified with $M_h/2$ and for the latter $\mu_s=M_h/2$ and
$\mu_h=M_t$ has been chosen.

One observes that the difference between SM and MSSM becomes small for
increasing $\tilde{m}_t$ which is expected since in this limit the spectrum
becomes heavy.  However, for smaller values of $\tilde{m}_t$ a
sizeable effect of the generic supersymmetric contribution is visible. For example, for
$\tilde{m}_t=400$~GeV a reduction of the SM cross section of about 5\% is
observed when including NNLO supersymmetric corrections.

The difference between the dashed and solid line in
Fig.~\ref{fig::sig_mssm}(a) quantifies the effect of the resummation of $\ln
(M_h^2/M^2_{\rm heavy})$ where $M_{\rm heavy}$ is a heavy mass scale present
in the calculation of $C_1$. It is negligible for large $\tilde{m}_t$,
however, for smaller values it can lead to a visible effect.

It is interesting to note that supersymmetric three-loop corrections to $C_1$
computed in Refs.~\cite{Pak:2012xr,Kurz:2012ff} provide an important
contribution to the difference of the solid and dotted curve in
Fig.~\ref{fig::sig_mssm}(a). In fact, if we choose $\tilde{m}_t=400$~GeV and
identify the three-loop matching coefficient in Eq.~(\ref{eq::leff}) with the
SM one a reduction of the cross section of only 3\% and not 5\% is observed.

Let us finally present results for $\sigma(pp\to h+X)$ which include in
addition bottom quark contributions up to NLO and furthermore also electroweak
corrections.  In Fig.~\ref{fig::sig_mssm}(b) we show the dependence on
$\tilde{m}_t$ at LO, NLO and NNLO (from bottom to top).  The dotted curves in
Fig.~\ref{fig::sig_mssm}(b) correspond to the solid ones of
Fig.~\ref{fig::sig_mssm}(a), i.e. they only include the top-sector
contribution. The inclusion of the bottom quark effects at NLO
(cf. Eq.~(\ref{eq::sigma})) leads to a reduction of about 5\% as shown by the
dashed curves. The reduction is basically independent of $\tilde{m}_t$ and
$\tan\beta$.\footnote{The dependence on $\tan\beta$ is studied in
  Ref.~\cite{Pak:2012xr}.}  Thus, even for $\tan\beta=20$ the bottom quark
effects are small for the considered scenarios and, hence, at NNLO the
approximation of vanishing bottom Yukawa coupling
is justified.  The reduction due to bottom quark effects
is to a large extent compensated by the electroweak corrections taken into
account multiplicatively as can be seen by the solid line which includes all
contributions of Eq.~(\ref{eq::sigma}).

To conclude this subsection let us remark that in the recent years
considerable progress in the computation of higher order supersymmetric
corrections to the Higgs boson production cross section has been achieved.
The supersymmetric NNLO corrections can affect the production cross section by
a few percent in case there is a splitting in the top squark masses by a few
hundred GeV and the overall scale of the spectrum is not too heavy. Such
effects are certainly relevant once the experimental precision for the cross
section has been reduced, in particular, once there
are hints for new particles from direct searches.




\section{\label{sec::beta}Renormormalization group functions in the SM to three
loops}

Renormalization group functions are fundamental quantities of each
quantum field theory.  In general, beta functions provide insights in
the energy dependence of cross sections, hints to phase transitions
and can provide evidence to the energy range in which a particular
theory is valid.  In the recent years the beta functions of all SM
couplings have been extended to three loops. This was partly triggered
by the discovery of a Higgs boson at the
LHC~\cite{Aad:2012tfa,Chatrchyan:2012ufa}. A precise running of
the Higgs boson self coupling from low energies to energies of the
order of the grand unification scale (GUT scale) is mandatory in order
to make firm statements about the stability of the Higgs boson vacuum.
A further motivation is connected to {the possibility of} 
gauge and/or Yukawa coupling
unification at high energies. Again, the renormalization group
functions are needed in order to transfer the knowledge of the
couplings at the electroweak scale to the high energy scales where
unification is expected.

Within the CRC/TR~9 several important contributions have been achieved
which are summarized in the following subsections. In particular, in
Subsections~\ref{sub::beta_def}--\ref{sub::num} we describe the
calculation of the SM beta functions and review the important
contributions to the topic. In Subsection~\ref{sub::susy} the
calculation of the three-loop SUSY QCD beta function is mentioned. In
Sections~\ref{sub::dec}--\ref{sub::dec_su5} decoupling constants,
which establish the relations between couplings and parameters in the
full and effective theories, are discussed. 
Furthermore, three applications, both of the decoupling
constants and the renormalization group functions, are presented in
Sections~\ref{sub::LET},~\ref{sub::su5} and~\ref{sub::Hpot}.
Further details can also be found in Ref.~\cite{Mihaila:2013wma}.


\subsection{\label{sub::beta_def}SM couplings and definition of the $\beta$ functions}

The SM is a product of the three subgroups ${\rm U}_Y(1)$, ${\rm SU}(2)_L$ and ${\rm
  SU}(3)_C$. To each one a coupling constant is assigned which is usually
denoted by $g_1$, $g_2$ and $g_3$, the gauge couplings.  Often it is
convenient to introduce
\begin{eqnarray}
  \alpha_i &=& \frac{g_i^2}{4\pi}\,, \quad i=1,2,3\,,
\end{eqnarray}
which will be used below in the expressions for the beta functions.
$\alpha_1$, $\alpha_2$ and $\alpha_3$ obey the following 
all-order relations to the fine structure constant $\alpha_{\rm QED}$,
the weak mixing angle $\theta_W$ and the strong coupling constant, which are
usually used in the SM
\begin{align}
  \alpha_1 &= \frac{5}{3}\frac{\alpha_{\rm QED}}{\cos^2\theta_W}\,,\notag\\
  \alpha_2 &= \frac{\alpha_{\rm QED}}{\sin^2\theta_W}\,,\notag\\
  \alpha_3 &= \alpha_s\,.
  \label{eq::alpha_123}
\end{align}
Note that in Eq.~(\ref{eq::alpha_123}) SU(5) normalization has been used which
leads to the factor 5/3 in the equation for $\alpha_1$.
Equation~(\ref{eq::alpha_123}) can as well be considered as a definition
for $\alpha_{\rm QED}$ and $\theta_W$.

For each massive fermion there exists a Yukawa coupling to the Higgs boson.
To lowest order it is given by
\begin{equation}\label{eq::def Yukawa}
  \alpha_x = \frac{\alpha_{\rm QED} m_x^2}{2 \sin^2\theta_W M_W^2}
  {\quad \mbox{with}\quad} x=e,\mu,\tau,u,d,c,s,t,b\,,
\end{equation}
where $m_x$ and $M_W$ are the fermion and W boson mass, respectively.

Finally, there is the Higgs boson self-coupling $\hat{\lambda}$, which
we define via the following term in the Lagrange density
\begin{eqnarray}
  {\cal L}_{\rm SM} &=& \ldots - (4\pi\hat{\lambda})(H^{\dagger}H)^2 + \ldots
  \,,
\end{eqnarray}
describing the quartic Higgs boson self-interaction.
$H$ is the Higgs doublet field in the SM.

Note that the Yukawa couplings of the lighter fermions are phenomenologically
irrelevant. For this reason we only keep $\alpha_t$, $\alpha_b$ and
$\alpha_\tau$ different from zero and define $\alpha_4=\alpha_t$,
$\alpha_5=\alpha_b$, $\alpha_6=\alpha_{\tau}$ and $\alpha_7=\hat{\lambda}$.
Some of the formulae presented below can easily be extended to the more
general case in an obvious way. 

Throughout this section we adopt the modified minimal subtraction (\msbar)
renormalization scheme. Note that in this scheme the beta functions are mass
independent which allows us to perform the calculations in the unbroken phase
of the SM where all particles are massless.

We define the beta functions as
\begin{equation}
  \mu^2\frac{d}{d\mu^2}\frac{\alpha_i}{\pi}=\beta_i(\{\alpha_j\},\epsilon)\,,
  \label{eq::beta_fc}
\end{equation}
where $\epsilon=(4-d)/2$ is the regulator of Dimensional Regularization  with
$d$ being the space-time dimension used for the evaluation of the momentum
integrals.  The dependence of the couplings $\alpha_i$ on the renormalization
scale is suppressed in the above equation.

The beta functions are obtained by calculating the renormalization constants
relating bare and renormalized couplings which we define via\footnote{Note that, in
  the case of $\alpha_7=\hat\lambda$ we have that $Z_{\alpha_7}$ contains
  terms proportional to $1/\hat\lambda$. The developed formalism, in
  particular Eq.~(\ref{eq::renconst_beta}) is nevertheless applicable.}
\begin{eqnarray}
  \alpha_i^{\text{bare}} &=&
  \mu^{2\epsilon}Z_{\alpha_i}(\{\alpha_j\},\epsilon)\alpha_i
  \,.
  \label{eq::alpha_bare}
\end{eqnarray}
where $i=1,\ldots,7$.
Taking into account that $\alpha_i^{\text{bare}}$ does
not depend on $\mu$, Eqs.~(\ref{eq::beta_fc}) and (\ref{eq::alpha_bare}) lead
to
\begin{eqnarray}
  \label{eq::renconst_beta}
  \beta_i = 
  -\left[\epsilon\frac{\alpha_i}{\pi}
    +\frac{\alpha_i}{Z_{\alpha_i}}
    \sum_{{j=1},{j \neq i}}^7
    \frac{\partial Z_{\alpha_i}}{\partial \alpha_j}\beta_j\right]
  \left(1+\frac{\alpha_i}{Z_{\alpha_i}}
    \frac{\partial Z_{\alpha_i}}{\partial \alpha_i}\right)^{-1}
  \,,
\mbox{}\hspace*{-5em}
  \nonumber\\
\end{eqnarray}
where the first term in the first factor of Eq.~(\ref{eq::renconst_beta}) 
originates from the term $\mu^{2\epsilon}$ in Eq.~(\ref{eq::alpha_bare}) 
and vanishes in four space-time dimensions. 
The second term in the first factor
contains the beta functions of the remaining six couplings of the SM. 

From the second factor of
Eq.~(\ref{eq::renconst_beta}) it is obvious that three-loop corrections to
$Z_{\alpha_i}$ are required for the computation of $\beta_i$ to the same loop
order.

Note that for the gauge couplings $\alpha_1$, $\alpha_2$ and $\alpha_3$ the
one-loop term of $Z_{\alpha_i}$ only contains $\alpha_i$, whereas at two loops
all couplings are present, except $\hat{\lambda}$. The latter appears for the
first time at three-loop level. As a consequence, for the three-loop
calculation of $\beta_1$, $\beta_2$ and $\beta_3$, it is necessary to know
$\beta_j$ for $j=4,5,6$ to one-loop order and only the $\epsilon$-dependent
term for $\beta_7$, namely $\beta_7 = - \epsilon \alpha_7/\pi$.  

This is different for the Yukawa couplings. Here, in the one-loop corrections
to the renormalization constants $\alpha_1, \ldots, \alpha_6$ 
appear which means that the two-loop gauge coupling beta functions
are needed in order to compute the three-loop term to the 
Yukawa beta function. $\hat{\lambda}$ is present in the Yukawa coupling
renormalization constant starting from two loops and thus the one-loop
term of $\beta_7$ is needed.

For the calculation of $\beta_7\equiv\beta_{\hat\lambda}$ all beta functions
except $\beta_3$ are required to two-loop accuracy since the corresponding
couplings are already present in the one-loop renormalization constant. The
strong coupling only enters at two loops and, hence, for $\beta_3$ only the
one-loop result is needed.

Before presenting some details on the calculation of three-loop corrections to
the renormalization constants in the next subsection we want to end this
subsection with a summary of important milestones for the calculation of the
beta functions in the SM:
\begin{itemize}
\item The one-loop beta functions in gauge theories along with the discovery
  of asymptotic freedom have been presented in
  Refs.~\cite{Gross:1973id,Politzer:1973fx}.
\item The corresponding two-loop corrections 
  \begin{itemize}
  \item in gauge theories without fermions~\cite{Jones:1974mm,Tarasov:1976ef},
  \item in gauge theories with fermions neglecting Yukawa
    couplings~\cite{Caswell:1974gg,Egorian:1978zx,Jones:1981we},
  \item involving Yukawa corrections~\cite{Fischler:1981is},
  \end{itemize}
  are also available.
\item The two-loop gauge coupling beta functions in an arbitrary quantum field
  theory have been considered in Refs.~\cite{Machacek:1983tz,Jack:1984vj}.
\item Two-loop corrections to the renormalization group functions for the 
  Yukawa and Higgs boson self-couplings in the SM are also
  known~\cite{Fischler:1982du,Machacek:1983fi,Machacek:1984zw,Jack:1984vj,Ford:1992pn,Luo:2002ey}.
\item The contribution of the scalar self-interaction at three-loop
  order has been computed in~\cite{Curtright:1979mg,Jones:1980fx}.
\item The gauge coupling beta function in QCD 
  to three loops is known from Refs.~\cite{Tarasov:1980au,Larin:1993tp}.
\item The three-loop corrections to the gauge coupling beta function 
  involving two strong and one top quark Yukawa coupling
  have been computed in Ref.~\cite{Steinhauser:1998cm}.
\item The three-loop corrections for a general quantum field theory based on a
  single gauge group have been computed in~\cite{Pickering:2001aq}.
\item The complete three-loop corrections to the gauge coupling 
  beta functions in the SM has been computed in
  Refs.~\cite{Mihaila:2012fm,Mihaila:2012pz,Bednyakov:2012rb} 
\item The four-loop corrections in QCD 
  are known from Refs.~\cite{vanRitbergen:1997va,Czakon:2004bu}.
\item The dominant three-loop corrections to the renormalization group
  functions of the top quark Yukawa and the Higgs boson self-coupling have been
  computed in Ref.~\cite{Chetyrkin:2012rz}.
  In that calculation the gauge
  couplings and all the Yukawa couplings except the one of the top quark have
  been set to zero.
\item Complete three-loop corrections to the Yukawa coupling beta
  functions in the SM has been computed in
  Ref.~\cite{Bednyakov:2012en} (see also Ref.~\cite{Bednyakov:2014pia}). 
\item Complete three-loop corrections to the Higgs
  boson self-coupling beta function in the SM has been computed in
  Refs.~\cite{Chetyrkin:2013wya,Bednyakov:2013eba}, even with complex Yukawa
  matrices~\cite{Bednyakov:2013cpa}.
\end{itemize}


\subsection{\label{sub::calc}Calculation of the renormalization constants to three loops}

In order to compute the renormalization constant of a coupling one has to
consider loop corrections to a vertex involving this coupling. In addition the
wave function renormalization constants for the external particles have to be
computed, see, e.g., Refs.~\cite{Steinhauser:1998cm,Steinhauser:2002rq}.
For example, if we consider the $N$-point
vertex with external fields $\phi_1,\ldots,\phi_n$ and denote its
coupling constant by $g$, we obtain
\begin{eqnarray}
  Z_g &=& \frac{Z_{\phi_1\cdots\phi_N}}{\sqrt{Z_{\phi_1}\cdots Z_{\phi_N}}}\,,
  \label{eq::ZZZ}
\end{eqnarray}
where the $Z_{\phi_i}$ are the wave function renormalization constants for the
$\phi_i$, $Z_{\phi_1\cdots\phi_N}$ is the corresponding vertex renormalization
constant, and $Z_g$ the renormalization constant for the coupling $g$.  
Formulae can be derived where the $\overline{\rm MS}$ $Z$ factors are obtained
from the ultraviolet-divergent part of amputated Green's functions 
(accompanied by higher order $\epsilon$ terms of lower-order contributions;
see, e.g., Ref.~\cite{Steinhauser:2002rq}).

\begin{figure}[t]
  \begin{center}
    \includegraphics[width=\linewidth]{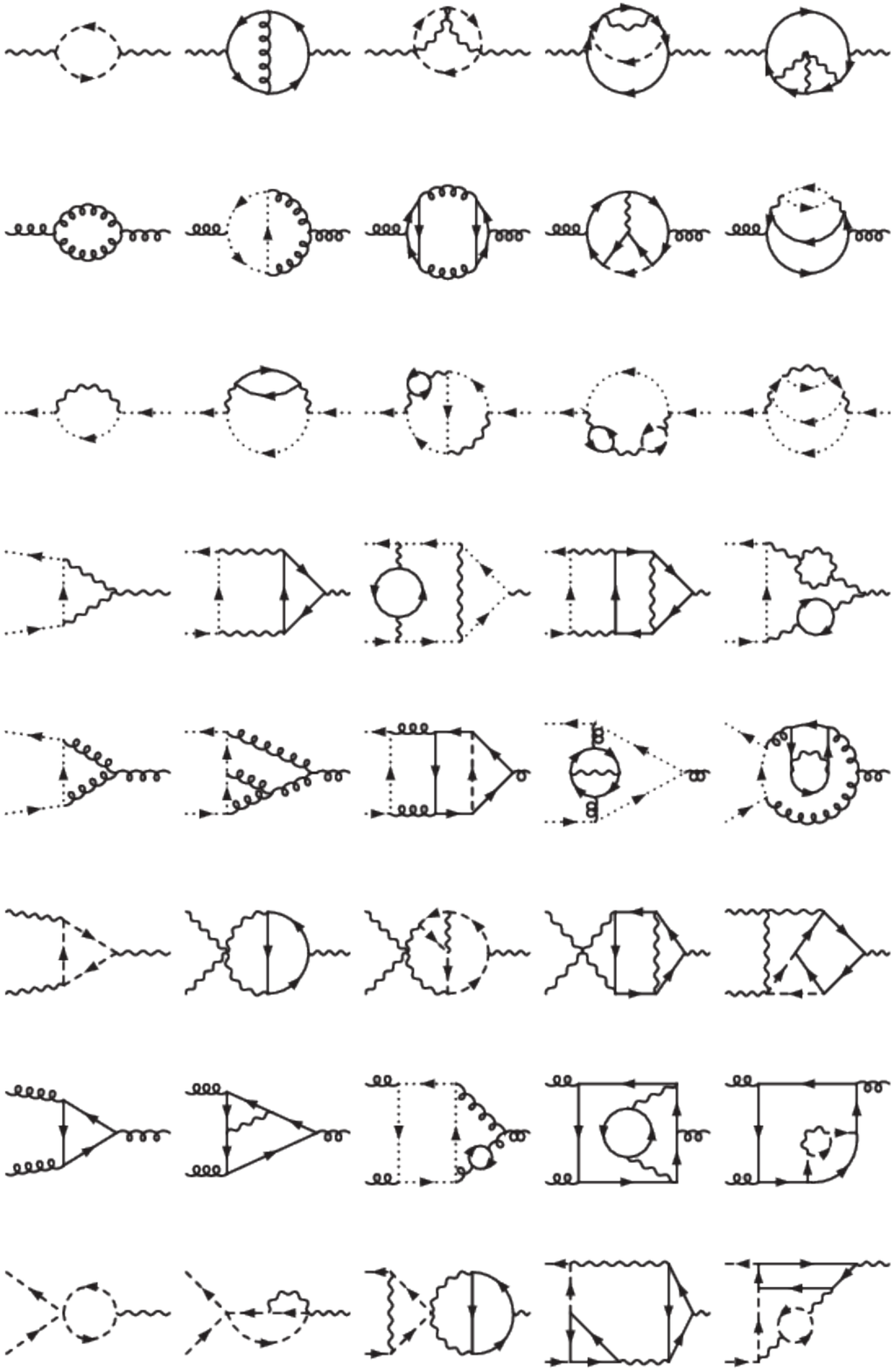}
    \caption{\label{fig::diags}
      Sample Feynman diagrams contributing to the
      Green's functions which can be used for the calculation of the gauge
      coupling renormalization constants.  Solid, dashed, dotted, curly and
      wavy lines correspond to fermions, Higgs bosons, ghosts, gluons and
      electroweak gauge bosons, respectively.}
  \end{center}
\end{figure}

Sample Feynman diagrams for
the case of the gauge couplings are shown in Fig.~\ref{fig::diags}. The
renormalization constants for the couplings $g_1$ and $g_2$ can, e.g., be
computed from the gauge boson two-point function (first line), ghost two-point
function (third line) and the gauge boson-ghost vertex (fourth line).
In the case of the Yukawa coupling the Higgs boson-fermion
vertex can be used (together with the corresponding two-point functions)
and for the Higgs boson self coupling the vertices involving four
scalar particles.

In the practical evaluation of the loop integrals one exploits the fact that
in the $\overline{\rm MS}$ scheme the renormalization constants do not depend
on the kinematical quantities like external momenta or particle masses. Thus,
it is possible to choose convenient configurations which allow for a simple
evaluation of the integrals. In case two- or three-point Green's functions
have to be computed it is convenient to treat all involved particles as
massless and keep only one external momentum non-zero. This leads to massless
propagator-type integrals which up to three loops can be computed with the
help of the {\tt FORM} program {\tt MINCER}~\cite{Larin:1991fz}. This approach
can in principle lead to infrared divergences. However, introducing a small
mass as potential infrared regulator in combination with asymptotic expansion
it is straightforward to check infrared safety, see
Ref.~\cite{Mihaila:2012pz}. This method has been used to compute the gauge and
Yukawa coupling renormalization constants up to three loops.

In case the renormalization constants have to be extracted from four-point
functions, like the renormalization constant of the Higgs boson self coupling,
the described procedure cannot be applied since nullifying all but one
external momenta inevitably leads to infrared divergences.  Thus, it is more
profitable to set all external momenta to zero and introduce a common mass $M$
to all particles. Up to three loops the resulting loop
integrals are well studied in the literature and a automated calculation is
possible with the help of the {\tt FORM} program {\tt
  MATAD}~\cite{Steinhauser:2000ry}. A minor disadvantage of this approach is
that additional non-standard counterterm contributions have to be introduced
for the mass parameter $M$. A details description of this method can be found
in in Ref.~\cite{Chetyrkin:1997fm}. It has been used to compute three-loop
corrections to the anomalous dimension matrix necessary for analyzing the
decay $\bar{B}\to X_s\gamma$ at NNLO~\cite{Chetyrkin:1996vx}.

In general a large number of Feynman diagrams is involved in the calculation
of three-loop SM Green's functions ranging up to
a few millions for the Higgs boson four point functions. Thus, an automated
setup is mandatory. 

\begin{figure}[t]
  \includegraphics[width=\linewidth]{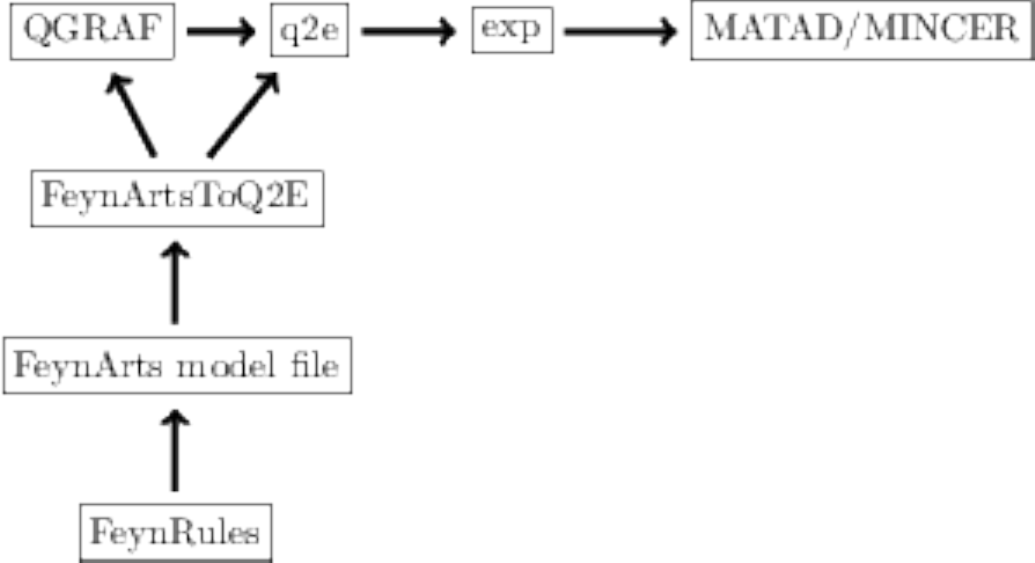}
  \caption{\label{fig::setup}Overview of our automated setup. 
      Calling up the programs in the
    uppermost line determines and evaluates a given process in a given
    model. The vertical workflow leads to the implementation of a new model in
    the setup.}
\end{figure}

In Refs.~\cite{Mihaila:2012fm,Mihaila:2012pz} a well-tested chain of programs
has been used that work hand-in-hand: \verb|QGRAF|~\cite{Nogueira:1991ex}
generates all contributing Feynman diagrams. The output is passed via
\verb|q2e|~\cite{Harlander:1997zb,Seidensticker:1999bb}, which transforms
Feynman diagrams into Feynman amplitudes, to
\verb|exp|~\cite{Harlander:1997zb,Seidensticker:1999bb} that generates
\verb|FORM|~\cite{Vermaseren:2000nd} code.  The latter is processed by
\verb|MINCER|~\cite{Larin:1991fz} and/or
\verb|MATAD|~\cite{Steinhauser:2000ry} that compute the Feynman integrals and
output the $\epsilon$ expansion of the result.  The parallelization of the
latter part is straightforward as the evaluation of each Feynman diagram
corresponds to an independent calculation.  The input for \verb|QGRAF| and
\verb|q2e| is provided by the program \verb|FeynArtsToQ2E| which translates
\verb|FeynArts|~\cite{Hahn:2000kx} model files into model files processable by
\verb|QGRAF| and \verb|q2e|.  Furthermore, it is possible to apply the package
\verb|FeynRules|~\cite{Christensen:2008py} in order to generate model files
for \verb|FeynArts|.  The complete workflow is illustrated in
Fig.~\ref{fig::setup}.

A similar level of automation has been obtained in
Refs.~\cite{Bednyakov:2012rb,Bednyakov:2012en} where 
\verb|LanHEP|~\cite{Semenov:2010qt}, 
\verb|FeynArts|~\cite{Hahn:2000kx}, 
\verb|MINCER|~\cite{Larin:1991fz},
\verb|color|~\cite{vanRitbergen:1998pn} and 
\verb|DIANA|~\cite{Tentyukov:1999is}
has been used.

In Ref.~\cite{Chetyrkin:2013wya} also {\tt QGRAF} has been used for
the diagram generation. The output is further processed with {\tt
  GEFICOM}~\cite{GEFICOM,Steinhauser:1998ry,GEFICOM2} and the
resulting three-loop integrals are computed with {\tt MINCER} and {\tt
  MATAD}.\footnote{Some of the program packages, which are not publicly
  available, can be obtained from the authors upon request; see also~\cite{sfb_software}.}

For some of the renormalization constants, in particular the ones related to
the Yukawa couplings, a careful treatment of the $\gamma_5$ matrix is
required. A practical prescription based on the work~\cite{Larin:1993tq} for the
computation of renormalization constants is given in
Ref.~\cite{Harlander:2009mn} and has been adopted in
Refs.~\cite{Bednyakov:2012rb,Bednyakov:2012en,Chetyrkin:2013wya}.


\subsection{\label{sub::num}Numerical results}

\begin{figure}[t]
  \begin{center}
  \includegraphics[width=0.9\linewidth]{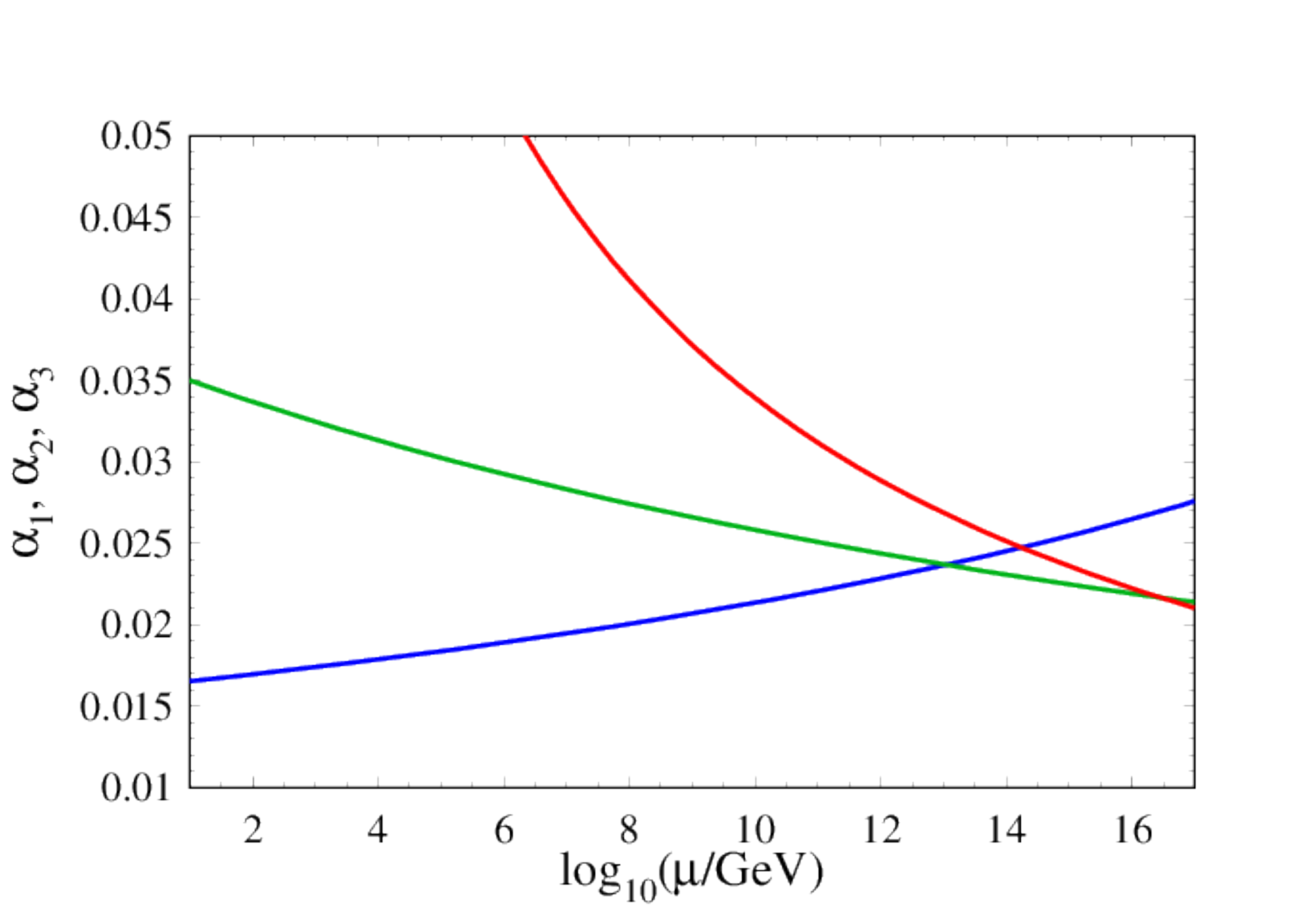}
  \\(a)\\
  \includegraphics[width=0.9\linewidth]{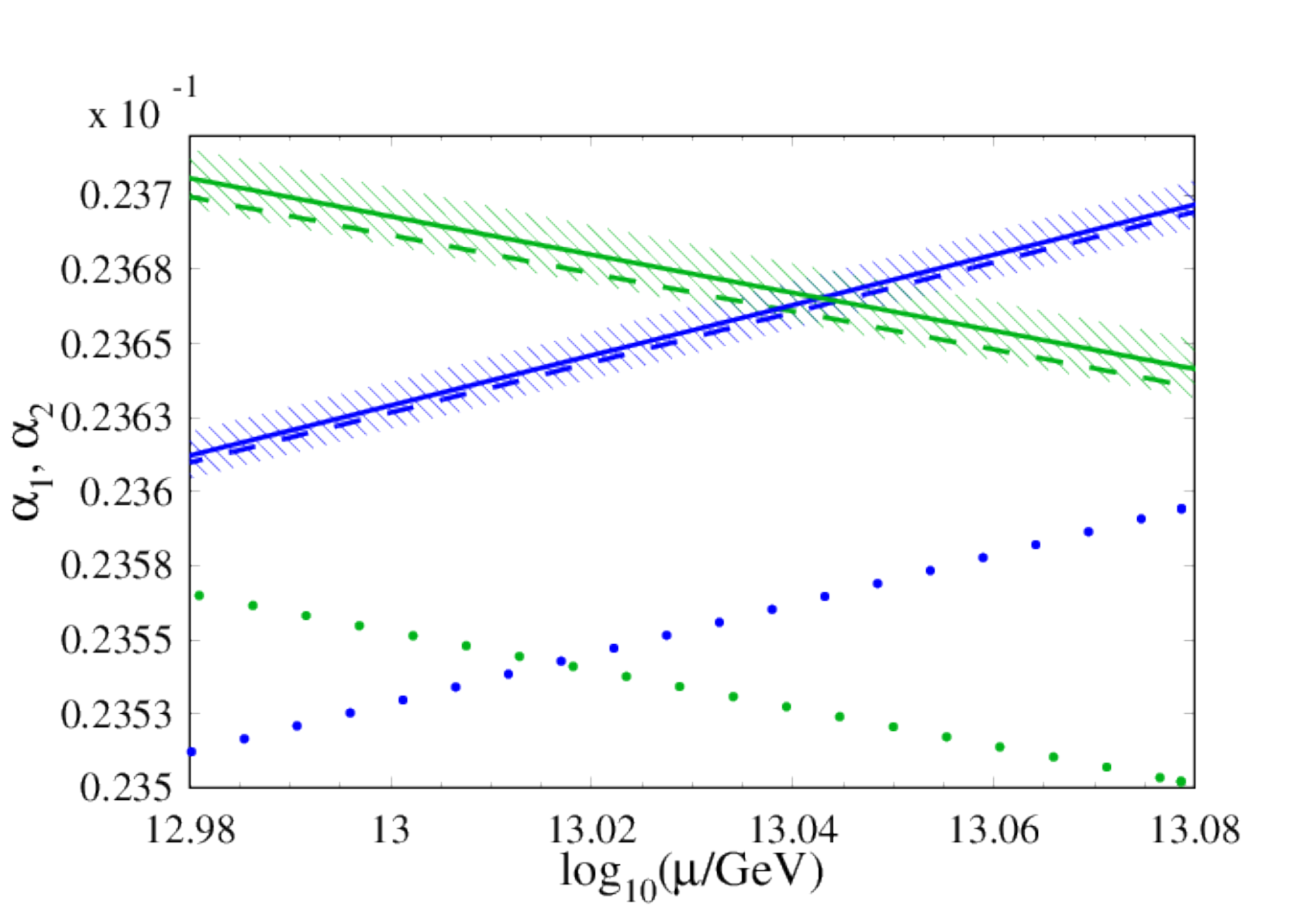}
  \\(b)
  \caption{\label{fig::gauge} (a) The running of the gauge couplings at three
    loops. The curve with the smallest initial value corresponds to $\alpha_1$
    (blue), the middle curve to $\alpha_2$ (green), and the curve with the
    highest initial value to $\alpha_3$ (red). (b) Magnification of the
    intersection region of $\alpha_1$ and $\alpha_2$ where the dotted, dashed
    and solid lines correspond to one-, two- and three-loop precision,
    respectively.  The bands around the three-loop curves visualize the
    experimental uncertainty. (Figure taken from Ref.~\cite{Mihaila:2012pz}.)}
  \end{center}
\end{figure}

We refrain from displaying analytic results for the beta functions which can
be found in the original publications.  Rather we briefly discuss the
numerical impact of the three-loop term.  In Fig.~\ref{fig::gauge} the running
of the three gauge couplings is shown assuming that the SM is valid up to high
scales. In panel~(a) the energy varies over 16 orders of magnitude. It is
interesting to have a closer look to the intersection region of $\alpha_1$ and
$\alpha_2$ which is shown in panel~(b) where the one-, two- and three-loop
results are shown as dotted, dashed and solid lines. The bands around the
three-loop curves reflect the numerical uncertainties for $\alpha_1(M_Z)$ and
$\alpha_2(M_Z)$ which are given by\footnote{See Ref.~\cite{Mihaila:2012pz} for
  more details.}
\begin{eqnarray}
  \label{eq::values}
  \alpha_1^{\msbarformel}\left(M_{{Z}}\right) &=& 0.0169225 \pm 0.0000039\,,\nonumber\\
  \alpha_2^{\msbarformel}\left(M_{{Z}}\right) &=& 0.033735 \pm 0.000020\,,
\end{eqnarray}
Defining the difference between successive loop orders as remaining
theoretical uncertainty one observes that without the three-loop corrections
the theory uncertainty is much larger than the experimental one. However,
after including the three-loop term the experimental error dominates over the
theoretical one.

\begin{figure}[t]
  \begin{center}
  \includegraphics[width=1.0\linewidth]{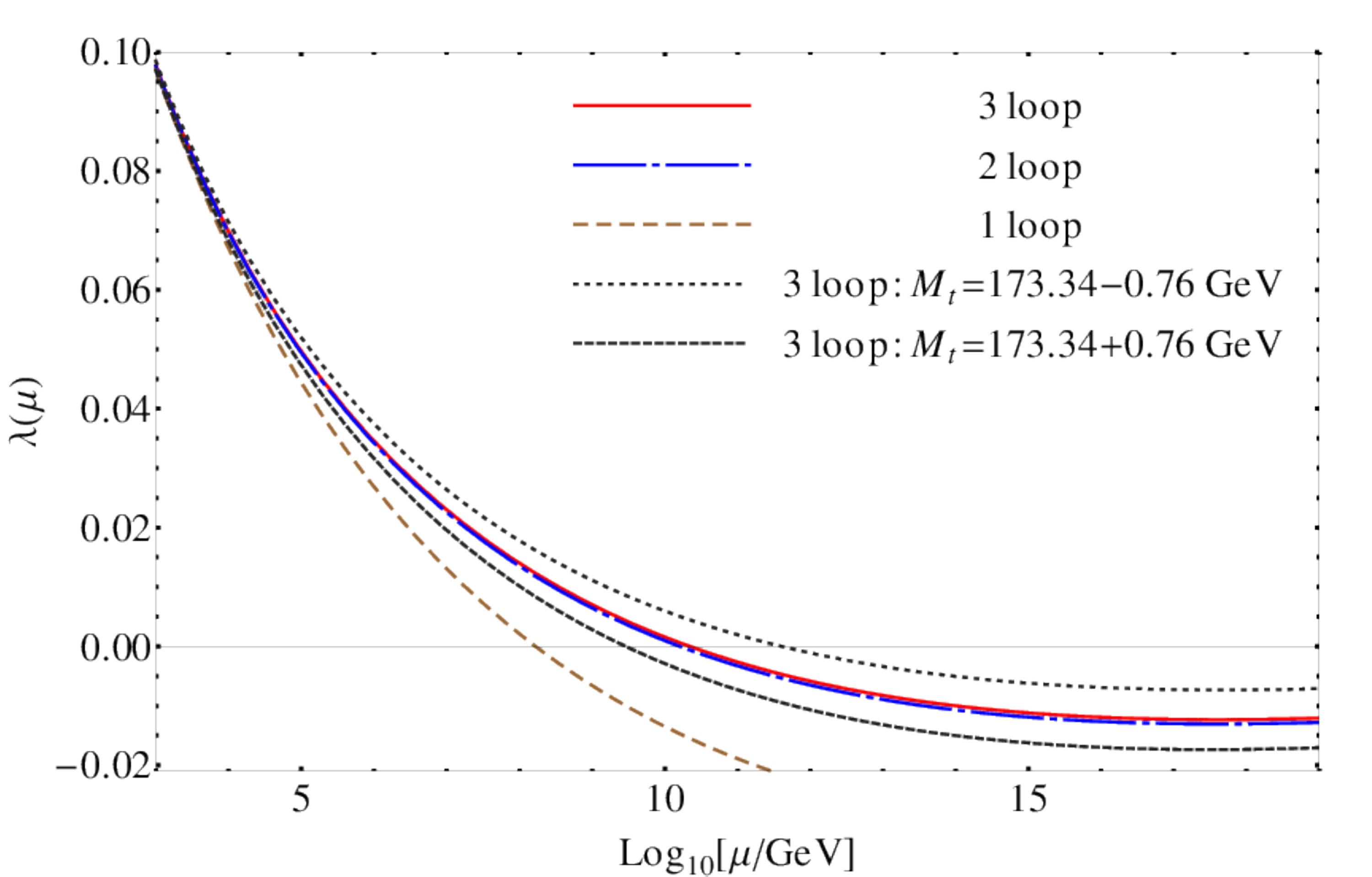}
  \caption{\label{fig::lambda} Evolution of $\lambda=4\pi\hat\lambda$ 
    with one- (brown, dashed), two- (blue, dash-dotted) and three-loop
    (red, solid) accuracy using initial conditions for $\mu=M_t$ (see
    Ref.~\cite{Zoller:2014xoa} for details). For comparison the uncertainty
    induced by the top quark pole mass is shown as (black) dotted
    lines. Figure taken from Ref.~\cite{Zoller:2014xoa}.}
  \end{center}
\end{figure}

As a further example we show in Fig.~\ref{fig::lambda} the running of the
quartic coupling up to the Planck scale with initial conditions taken for
$\mu=M_t$. One observes quite significant effects when going from one- to
two-loop accuracy. The three-loop corrections only lead to a small shift
indicating a stabilization of the perturbative expansion.  Notably, the
theory uncertainty due to the running
is now negligible as compared to the parametric one which
is dominated by the top quark mass.  In Fig.~\ref{fig::lambda} the
corresponding effect is shown as dotted lines.  From Fig.~\ref{fig::lambda}
one deduces that $\lambda$ becomes negative for $\ln(\mu/\mbox{GeV})\gsim
10.36$~\cite{Zoller:2014xoa}. Further
details can be found in Subsection~\ref{sub::Hpot}.


\subsection{\label{sub::susy}Beta function in the supersymmetric QCD}

Before we report on the diagrammatic calculation of the three-loop gauge beta
function in the supersymmetric QCD, few comments on the used regularisation
scheme are in order.

Whereas Dimensional Regularization in combination with the $\overline{\rm MS}$
scheme is the canonical choice for higher order calculations within QCD, it is
less appropiate for supersymmetric theories since it explicitely breaks
supersymmetry. This can easily be understood by counting degrees of freedom of
fermionic and spin-1 fields. Whereas the former has four degrees of freedom
the latter has $d$ degrees, where $d$ is the space-time dimension and
$d\not=4$.  Thus a modification of the theory is necessary to render the two
numbers equal. A convenient choice has been introduced in
Ref.~\cite{Siegel:1980qs}, so-called Dimensional Reduction.  The
essential difference between Dimensional Reduction
and Dimensional Regularization is that the
continuation from $4$ to $d$ dimensions is made by {\it compactification\/}
or {\it dimensional reduction}.  In this scheme, the momentum (or space-time)
integrals are $d$-dimensional in the usual way, whereas the number of field
components remains unchanged equal to four, and consequently supersymmetry is
undisturbed (see also Refs.~\cite{Capper:1979ns,Stockinger:2005gx}.) In
practice it is convenient to implement Dimensional Reduction
by introducing a new particle, the
so-called $\varepsilon$ scalar, that account for the additional $4-d$
components of the gauge boson fields. In this way, the well-established rules
for computing momentum intregrals in
Dimensional Regularization can also be applied to calculations performed in
Dimensional Reduction.

An explicit calculation of the three-loop beta function for supersymmetric QCD
has been performed in Ref.~\cite{Harlander:2009mn} confirming results which
have previously been available in the
literature~\cite{Jack:1996vg,Pickering:2001aq}.  The main purpose of
Ref.~\cite{Harlander:2009mn} was to perform a consistency check of Dimensional
Reduction as
regularisation scheme for supersymmetric theories at three-loop order. For
this goal, several renormalization constants in supersymmetric QCD has been
computed to three-loop accuracy.  It has been shown that the same beta
function is obtained from all three-particle vertices involving gluons,
gluinos and $\varepsilon$-scalars.  The results of
Ref.~\cite{Harlander:2009mn} explicitly demonstrates the consistency of
Dimensional Reduction
with supersymmetry and gauge invariance, an important pre-requisite for
(high-order) precision calculations in supersymmetric theories, like the
three-loop corrections to the Higgs boson mass, cf. Section.~\ref{sec:mh}.

As a by-product of the calculation in~\cite{Harlander:2009mn}, the predicted
relation between the gauge beta function and the gluino mass anomalous
dimension was verified to three-loop order. In addition, the three-loop
results for the quark mass anomalous dimension~\cite{Jack:1996qq} were
confirmed.

The use of Dimensional Reduction (in contrast to Dimensional Regularization)
is one of the technical challenges of the calculation. A further difficulty,
which is discussed in detail in Ref.~\cite{Harlander:2009mn}, is the treatment
of $\gamma_5$ which only plays a sub-leading role in the gauge coupling beta
functions of the SM. Another technical complication is connected
to the Majorana nature of the gluino which is not treated consistently in {\tt
  QGRAF}~\cite{Nogueira:1991ex} and thus a program, {\tt majoranas.pl}, has
been developed which implements the prescription of Ref.~\cite{Denner:1992vza}
and adjusts the output of {\tt QGRAF}.

Let us for completeness present the result for the supersymmetric beta
function up to three-loop order. Defining
\begin{eqnarray}
  \beta^{\rm SQCD}(\alpha_s) &=&
  -\sum_{n\geq 0} \left(\frac{\alpha_s}{\pi}\right)^{n+2}\beta_n^{\rm SQCD}\,,
\end{eqnarray}
we obtain for the first three coefficients
\begin{equation}
\begin{split}
   \beta_0^{\rm SQCD} &=
         \frac{3}{4} C_A -\frac{ 1}{2}T_f\,,\\
   \beta_1^{\rm SQCD} &=
        \frac{3}{8}\,C_A^2
       - T_f \left(  \frac{1}{2}\,C_F + \frac{1}{4} C_A \right)\,,\\
   \beta_2^{\rm SQCD} &=
        \frac{21}{64} C_A^3 + T_f^2 \left( \frac{3}{8}\,C_F + \frac{1}{16}\,C_A \right)
       \\
&+ T_f\left( \frac{1}{4}\,C_F^2 - \frac{13}{16}\,C_AC_F 
        - \frac{5}{16}\,C_A^2 \right)\,,
\end{split}
\end{equation}
where $C_F=(n_c^2-1)/(2n_c)$, $C_A=n_c$ are the quadratic Casimir invariants for
SU($n_c$), and $2T_f=n_f$ is the number of quark flavors (which is
equal to the number of squark flavors in supersymmetric QCD).

\subsection{\label{sub::dec}Decoupling of heavy particles}

The anomalous dimensions considered in the previous subsections depend on the
active degrees of freedom of the theory. 
In QCD, which for the following general discussion
shall be used as a sample theory, this dependence is simply due to $n_f$, the
number of quarks contributing to the running. In general, $n_f$ changes when
flavour thresholds are crossed while increasing or lowering the energy scale.
Note that it is not sufficient to simply raise or lower $n_f$ in the
coefficients of the anomalous dimensions
and continue the running of the corresponding parameters
using the new beta function,
but a careful construction of the low-energy effective theory is necessary.
In the case of QCD this construction is straightforward since the effective
theory with $n_f-1$ active quark flavours has the same structure, i.e.,
contains the same operators as the full theory. The fields, masses and
couplings in the two theories are related by so-called decoupling constants. A
detailed discussion and the explicit construction of the effective theory can
be found in Refs.~\cite{Chetyrkin:1997un,Steinhauser:2002rq}. In these
references formulae for the decoupling constants are provided which allow to
compute the $n$-loop decoupling constants from light-particle Green's functions
with vanishing external momenta. The formalism resembles the renormalization
procedure in the $\overline{\rm MS}$ scheme. However, in contrast to the
renormalization constants, the decoupling constants also contain finite (and
also higher order $\epsilon$) terms.

We want to remark that the necessity of introducing decoupling constants
originates from the use of a mass-independent (here $\overline{\rm MS}$)
renormalization scheme.  As a 
consequence the strong coupling constant $\alpha_s$ is not a physical quantity
in the sense that it is not defined via a Green's function which (at least in
principle) can be measured.  Furthermore, $\alpha_s(\mu)$ is not a continuous
function of $\mu$ but has finite steps at the energy scale where the heavy
quark is integrated out.  In Refs.~\cite{Jegerlehner:1998zg,Chetyrkin:2008jk}
a physical definition of the strong coupling has been introduced in the
so-called momentum subtraction (MOM) scheme and it has been shown that running
in this scheme, where the decoupling of heavy particles is automatic, is
equivalent to the running and decoupling procedure in the $\overline{\rm MS}$.
Let us in the following briefly comment on the comparison of the MOM and
$\overline{\rm MS}$ couplings before we continue our considerations in the
$\overline{\rm MS}$ scheme.

\begin{figure}[t]
  \centering
  \includegraphics[width=0.9\linewidth]{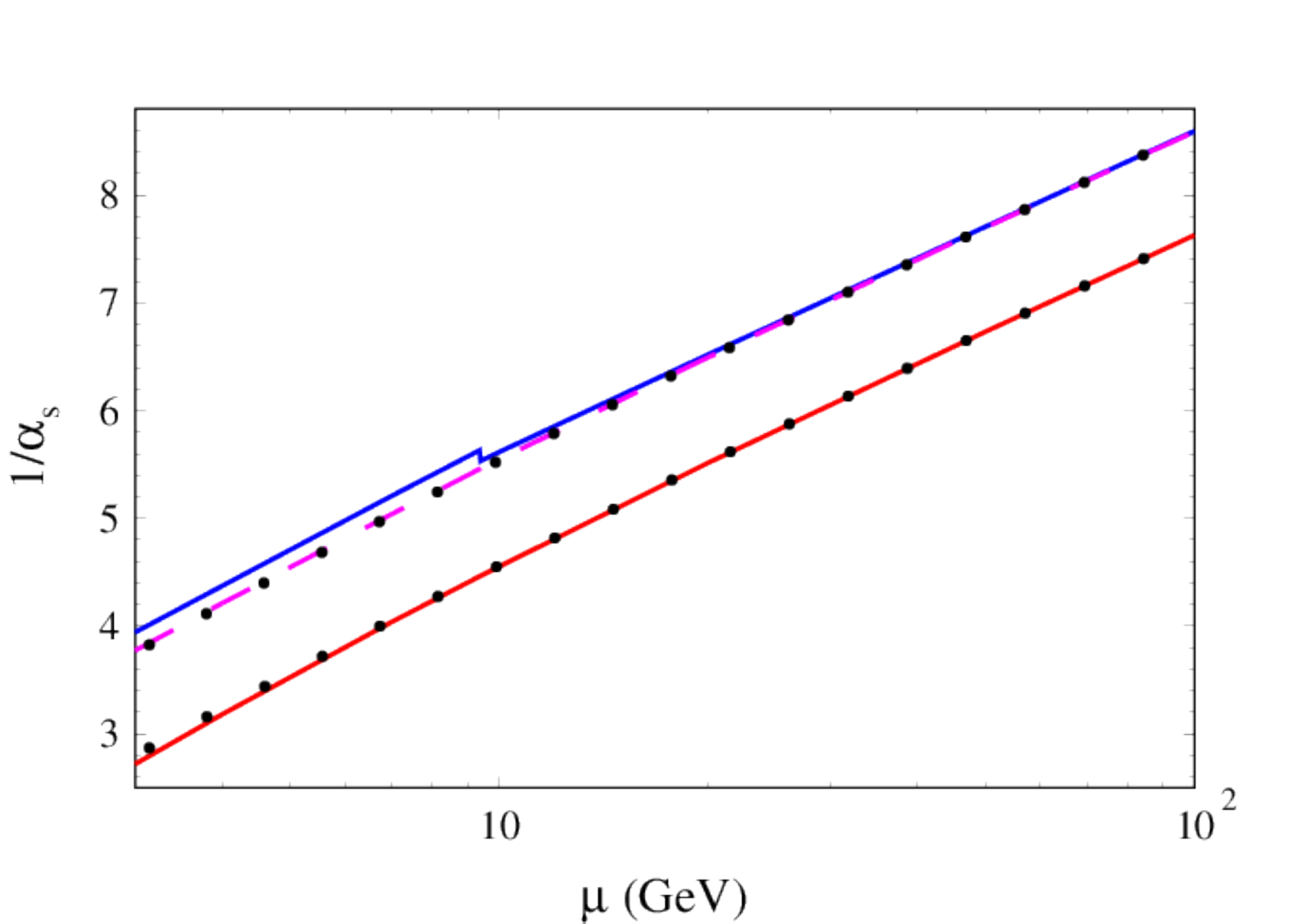}
  \caption[]{\label{fig::MOM_MS}$1/\alpha_s$ as a function of $\mu$.  The
    (blue) upper solid line containing a step for $\mu=2M_b$ corresponds to
    the $\overline{\rm MS}$ result and the (red) lower solid line to the
    result in the MOM scheme. The (pink) dashed line represents $1/\alpha_s$
    in the $\overline{\rm MOM}$ scheme.  The (black) dotted lines lying on top
    of the MOM and $\overline{\rm MOM}$ result correspond to the results
    obtained from $\overline{\rm MS}$ value of $\alpha_s$ using the proper
    conversion formulae (see Ref.~\cite{Chetyrkin:2008jk}).
    Figure taken from Ref.~\cite{Chetyrkin:2008jk}.}
\end{figure}

In Fig.~\ref{fig::MOM_MS} we show that the two schemes are equivalent by
plotting the inverse strong coupling as a function of $\mu$ both for the
$\overline{\rm MS}$, MOM and $\overline{\rm MOM}$\footnote{The $\overline{\rm
    MOM}$ scheme is a version of the MOM scheme which leads to numerical
  values for $\alpha_s$ close to the ones in the $\overline{\rm MS}$ scheme.}
schemes, where in all cases the three-loop approximation is used for the
running and the conversion between the schemes.  We choose
$\alpha_s^{(5)}(M_Z)$ as the input quantity and convert at this scale to the
other two schemes.  The evolution of the $\overline{\rm MS}$ coupling to lower
$\mu$ values is shown by the upper solid lines with a step at the value
$\mu=2M_b$ since at that scale the bottom quark is decoupled.  Numerically
very close is the dashed curve representing the $\overline{\rm MOM}$ scheme
result.  The lower solid line represents the result in the MOM scheme. Both
for the MOM and $\overline{\rm MOM}$ results, the conversion is performed for
$\mu=M_Z$, and the running to other values of $\mu$ is achieved using the
corresponding $\beta$ function. The dotted lines on top of the MOM and
$\overline{\rm MOM}$ curves represent the results where the transformation
from the $\overline{\rm MS}$ values is performed just at the considered value
of $\mu$. This shows that the $\overline{\rm MS}$ scheme including the
described decoupling procedure is equivalent to a physical scheme with a
physical definition of $\alpha_s$.

Let us now return to QCD with $\alpha_s$ defined in the $\overline{\rm MS}$
scheme.  The phenomenologically most important decoupling constants are the
ones of $\alpha_s$, $\zeta_{\alpha_s}$, and the light quark masses,
$\zeta_{m_q}$.  Two-loop corrections to $\zeta_{\alpha_s}$ have been computed
for the first time in Refs.~\cite{Bernreuther:1981sg,Larin:1994va}, however,
relatively complicated integrals had to be solved. For example, in
Ref.~\cite{Larin:1994va} a three-loop calculation for the $Z$ boson decay rate
has been performed. In Ref.~\cite{Chetyrkin:1997un} the two-loop result has
been checked and the three-loop results for $\zeta_{\alpha_s}$ and
$\zeta_{m_q}$ has been added. The latter required the computation of
three-loop one-scale vacuum integrals. The four-loop corrections to the
decoupling constants have been obtained by two independent calculations
performed in Refs.~\cite{Schroder:2005hy,Chetyrkin:2005ia}.

The loop-order used for the running and the one used for the decoupling are
related. In fact, $n$-loop running requires $(n-1)$-loop decoupling
relations. This can be seen by considering a physical quantity $R$ for which the
perturbative expansion is know up to order $\alpha_s^n$. Of course, in such a
situation one would apply $n$-loop corrections to the beta function. However,
using $n$-loop decoupling relations would affect the $\alpha_s^{(n+1)}$ term of
$R$ which is beyond the considered loop-order.

Let us in the following demonstrate the effect of higher order corrections to
the running and decoupling by considering the relation between
$\alpha_s^{(3)}(M_\tau)$ and $\alpha_s^{(5)}(M_Z)$ (see also
Refs.~\cite{Rodrigo:1997zd,Chetyrkin:1997un,Schroder:2005hy,Chetyrkin:2005ia}
where the charm and bottom flavour threshold is crossed at the scales $\mu_c$
and $\mu_b$, respectively. Note that these scales are not determined by
theory. On general grounds one expects that the result for
$\alpha_s^{(5)}(M_Z)$ gets more and more insensitive on the precise choice of
the decoupling scales when including higher order corrections.  The procedure
to compute $\alpha_s^{(5)}(M_Z)$ from $\alpha_s^{(3)}(M_\tau)$ is as follows.
In a first step we calculate $\alpha_s^{(3)}(\mu_c)$ by integrating the beta
function to $n$-loop order with the initial condition
$\alpha_s^{(3)}(M_\tau)=0.332$.  Afterwards $\alpha_s^{(4)}(\mu_c)$ is
obtained from $\alpha_s^{(3)}(\mu_c) = \alpha_s^{(4)}(\mu_c)\zeta_{\alpha_s}$
using the $(n-1)$-loop approximation for $\zeta_{\alpha_s}$.  Next $n$-loop
running is used to obtain $\alpha_s^{(4)}(\mu_b)$ and the decoupling procedure
is applied in analogy to the charm threshold to arrive at
$\alpha_s^{(5)}(\mu_b)$.  Finally, we compute $\alpha_s^{(5)}(M_Z)$ using
again the $n$-loop beta function. For the on-shell charm and bottom quark
masses we use $M_c=1.65$~GeV and $M_b=4.75$~GeV, respectively. These values
are obtained by using three-loop relations between the $\overline{\rm MS}$ and
on-shell quark masses~\cite{Chetyrkin:1999ys,Chetyrkin:1999qi,Melnikov:2000qh}.

\begin{figure}[t]
\begin{center}
  \includegraphics[width=\linewidth]{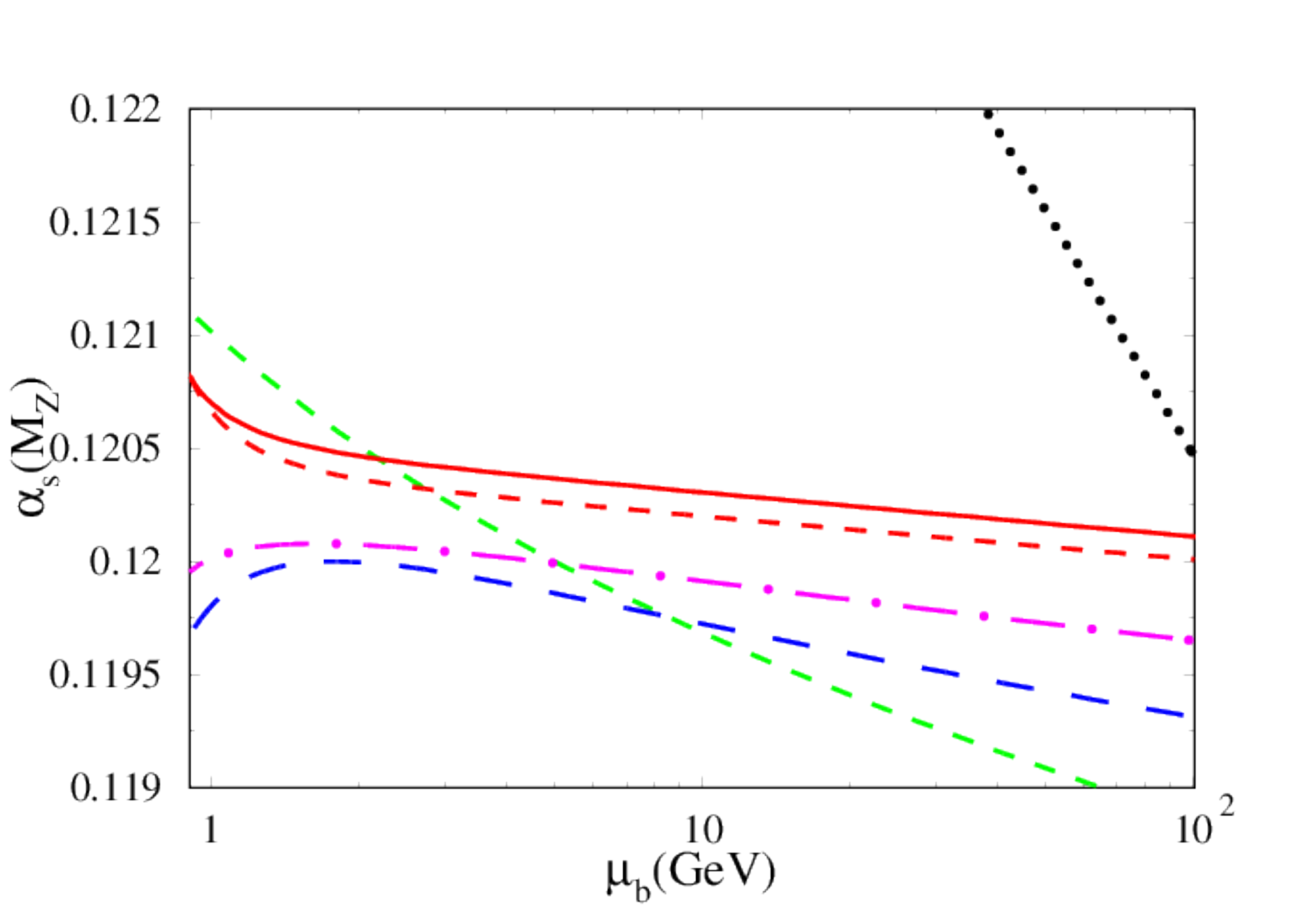}
  \caption{$\mu_b$ dependence of $\alpha_s^{(5)}(M_Z)$ calculated from
    $\alpha_s^{(4)}(M_\tau)$. The procedure is
    described in the text. The dotted, steep dashed, lower dashed,
    and dash-dotted line corresponds to 
    one- to four-loop running. The solid curve and the dashed line slightly
    below it includes the effect of the four-loop matching term.
  }
  \label{fig::asMZ}
\end{center}
\end{figure}

In Fig.~\ref{fig::asMZ} the result for $\alpha_s^{(5)}(M_Z)$ for fixed
$\mu_c=3$~GeV as a function of $\mu_b$ is displayed for the one- to five-loop
analysis.  For illustration, $\mu_b$ is varied rather extremely, by about two
orders of magnitude.  While the leading-order result (upper right dotted line)
exhibits a strong
logarithmic behaviour, the analysis is gradually getting more stable as we go
to higher orders.  The five-loop curve is almost flat for $\mu_b \ge 1$~GeV
(note the scale on the $y$ axis)
and demonstrates an even more stable behaviour than the four-loop analysis of
Ref.~\cite{Chetyrkin:1997un}. It should be noted that around $\mu_b \approx
1$~GeV both the three-, four- and five-loop curves show a strong variation
which can be interpreted as a sign for the breakdown of perturbation theory.
Besides the $\mu_b$ dependence of $\alpha_s^{(5)}(M_Z)$, also its absolute
normalization is significantly affected by the higher orders.  At the central
matching scale $\mu_b=M_b$, we encounter a rapid convergence behaviour.

Since the five-loop coefficient of the QCD beta function is not yet known we
choose the ($n_f$-independent) values $\beta_4^{(n_f)}=0$ (solid line) and
$\beta_4^{(n_f)}=150$ (dashed line parallel to the solid one)\footnote{The
  normalization corresponding to $\{\beta_0,\beta_1,\beta_2,\beta_3\} \approx
  \{1.92,2.42,2.83,18.85\}$ for $n_f=5$.}.  Larger
values would bring the five-loop curves even closer to the four-loop one.

From Fig.~\ref{fig::asMZ} it is possible to estimate a 
theory uncertainty on $\alpha_s^{(5)}(M_Z)$ as obtained from
$\alpha_s^{(3)}(M_\tau)$ due to missing higher order corrections. If
we restrict ourselves to a range of $\mu_b$ between 2~GeV and 10~GeV
and take the difference between the three- and four-loop curve as an
estimate for the uncertainty we obtain
$\delta\alpha_s^{(5)}(M_Z)\approx 0.0002$. The difference between the
four- and (dashed) five-loop curve would lead to
$\delta\alpha_s^{(5)}(M_Z)\approx 0.0003$. The variation of
$\alpha_s^{(5)}(M_Z)$ due to the variation of $\mu_b$ leads to an
additional uncertainty of $\delta\alpha_s^{(5)}(M_Z)\approx 0.0002$. A
similar uncertainty is obtained from the variation of $\mu_c$ between
2~GeV and 5~GeV.  (This can easily be checked with the program {\tt
  RunDec}~\cite{Chetyrkin:2000yt}.)  Thus a total uncertainty of
$\pm0.0004$ (obtained by adding the three uncertainties in quadrature)
should be assigned to $\alpha_s^{(5)}(M_Z)$. The uncertainties induced
by the errors in the quark masses are much smaller.

Note that it is straightforward to reproduce Fig.~\ref{fig::asMZ} using the
program {\tt RunDec}~\cite{Chetyrkin:2000yt} which is written in {\tt Mathematica}
or {\tt CRunDec}~\cite{Schmidt:2012az} written in {\tt C++}.

The formalism which has been described above allows the decoupling of one heavy
quark at a time. This procedure is certainly justified for the top quark being
more than factor 30 heavier than the bottom quark.  On the other hand, the ratio
between the bottom and charm quark mass is only approximately a factor
three. For this reason the simultaneous decoupling of the charm and bottom has
been studied in Ref.~\cite{Grozin:2011nk} and a decoupling constant relating the
strong coupling defined with three active flavours to the one in the
five-flavour theory has been derived.  These results can be used in order to
study the effect of power-suppressed terms in $M_c/M_b$ which are neglected in
the conventional approach~\cite{Chetyrkin:1997un}. Various analyses are
performed which indicate that the mass corrections present in the one-step
approach are small as compared to $\log(\mu^2/M_{c,b}^2)$ which are resummed
using the conventional two-step procedure, and, thus, even for the charm and
bottom quark case two-step decoupling is preferable.

\subsection{\label{sub::dec_susy}Decoupling at the SUSY scale}

As in QCD also in its supersymmetric extension, it is convenient to use a
mass-independent renormalization scheme.  Thus, by construction, the beta
function governing the running of $\alpha_s$ is independent of the particle
masses. It only depends on the particle content of the underlying theory,
i.e., the number of active quarks, squarks and gluinos.

In the MSSM, the one-loop decoupling relation for $\alpha_s$ has been computed
in Refs.~\cite{Hall:1980kf,Harlander:2004tp} and the two-loop relation is
known from Refs.~\cite{Harlander:2005wm,Bednyakov:2007vm,Bauer:2008bj} (see
also Ref.~\cite{Harlander:2007wh}). Due to many different mass scales exact
results cannot be obtained at three loops. Thus,
in Ref.~\cite{Kurz:2012ff} three different hierarchies
among the supersymmetric particle masses have been assumed to compute
the decoupling relation.

\begin{figure}[t]
  \centering
  \includegraphics[width=1.\linewidth]{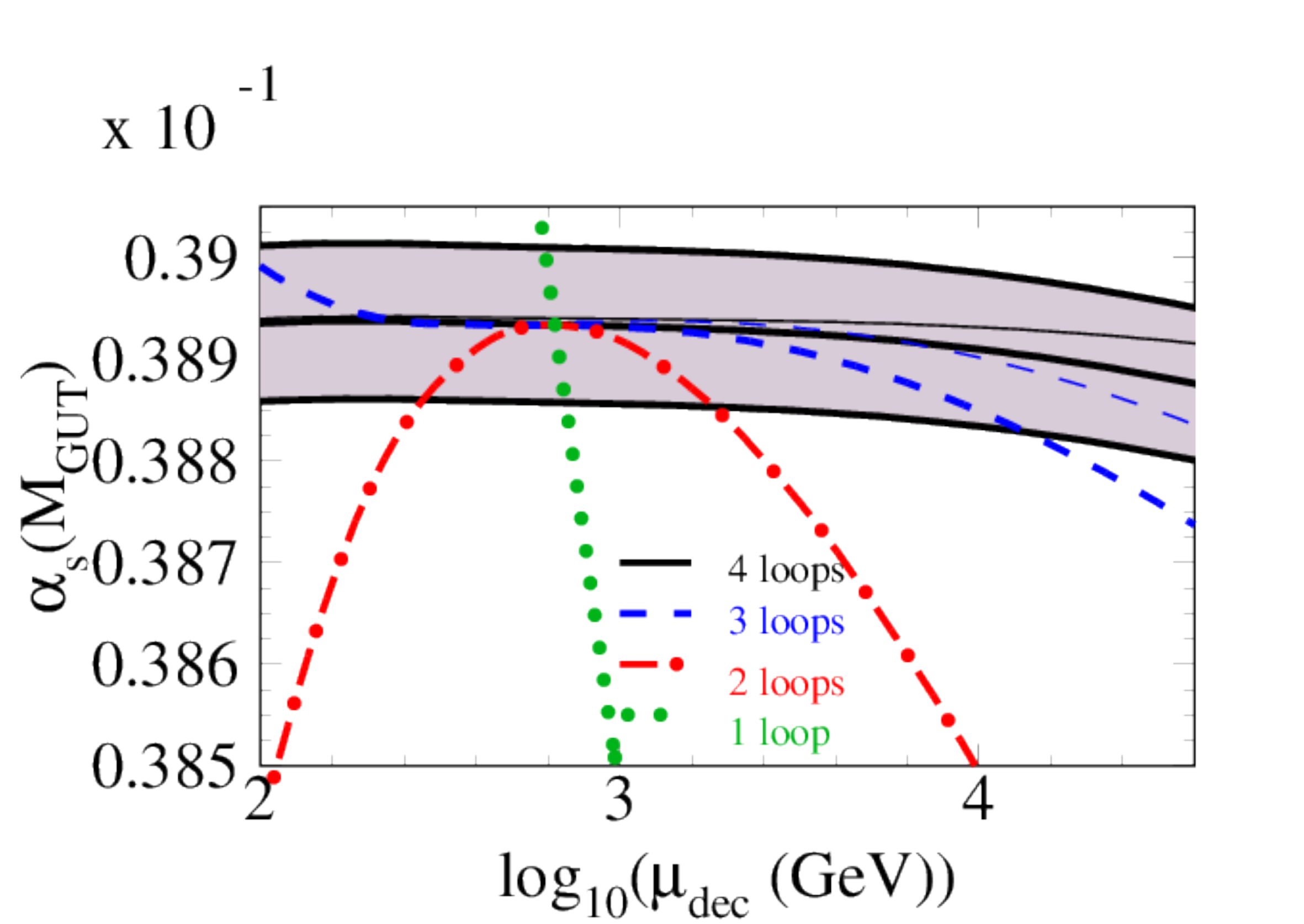}
  \caption[]{\label{fig::asGUT}$\alpha_s^{\rm (SQCD)}(M_{\rm GUT})$ as a function
    of $\mu_{\rm dec}$. Thick and thin 
    lines correspond to the one- and two-step scenario, respectively (see text).
    Thin lines are only shown for three- and four-loop running.
    (Figure taken from Ref.~\cite{Kurz:2012ff}.)}
\end{figure}

Let us in the following demonstrate the numerical effect of higher order
corrections to the decoupling constants by repeating the considerations of
Subsection~\ref{sub::dec}. However, instead of considering the strong coupling
with three and five active flavours we study the relation between
$\alpha_s^{\rm (SQCD)}(M_{\rm GUT})$ (defined in the $\overline{\rm DR}$
scheme) with $M_{\rm GUT}=2\cdot10^{16}$~GeV and $\alpha_s^{(5)}(M_Z)$ (in the
$\overline{\rm MS}$ scheme) as a function of the decoupling scale $\mu_{\rm
  dec}$. At this scale all supersymmetric particles are integrated out and the
transition from the MSSM to the SM is made.  We proceed as follows: In a first
step we run in the SM from $\mu=M_Z$ to $\mu=\mu_{\rm dec}$ where the
decoupling of the top quark and the supersymmetric particles is performed
simultaneously and $\alpha_s^{(5)}(\mu_{\rm dec})$ is transformed to
$\alpha_s^{(\rm SQCD)}(\mu_{\rm dec})$.  The use of the SUSY QCD $\beta$
function finally leads to $\alpha_s^{\rm (SQCD)}(M_{\rm GUT})$.  The thick
lines in Fig.~\ref{fig::asGUT} correspond to this procedure where dotted,
dash-dotted, dashed and solid lines correspond to one-, two-, three- and
four-loop running and the use of decoupling constants to one order less.  
For a complete list of input parameters we refer to Ref.~\cite{Kurz:2012ff}.

As expected, the inclusion of higher order corrections reduces dramatically
the dependence on $\mu_{\rm dec}$ leading to an almost flat behaviour at four
loops, even though the decoupling scale is varied over more than two orders of
magnitude. The band around the four-loop curve reflects the uncertainty of the
strong coupling constant which has been chosen as $\alpha_s^{(5)}(M_Z) =
0.1184 \pm 0.0007$.  It is interesting to note that $\mu_{\rm dec}=M_Z$ is in
general a less favourable choice unless three- or four-loop running is used.
On the contrary, for $\mu_{\rm dec}\approx1$~TeV, which approximatley
corresponds to the masses of the supersymmetric particles, higher order
corrections are quite small.

Alternatively, in order to obtain the thin lines we integrate out the top
quark in a separate step at the scale $\mu=M_t$ ($M_t$ is the on-shell
top quark mass) and transform afterwards $\alpha_s^{(6)}(M_t)$ to
$\alpha_s^{(\rm SQCD)}(M_{\rm GUT})$ in analogy to the previous description.
The three- and four-loop curves show in this variant an even flatter
behaviour.

\subsection{\label{sub::dec_su5}Decoupling at the GUT scale}

Another type of threshold corrections that have to be considered in the
context of GUTs are those generated by the super-heavy
particles present in such models. In particular, the  gauge coupling
unification (one of the most important predictions of a GUT)
is very sensitive to these corrections. This property enables us to constrain
the models, once the low energy values of the gauge couplings and their 
evolution  are known precisely. In the following, we restrict our discussion
 to the effects of the threshold corrections at the  super-heavy scale on the energy evolution
of the SM gauge couplings. 

 The new features of the GUT threshold
corrections as compared with those discussed in the previous subsections are
related to the spontaneous breaking of the gauge symmetry. Explicitly, they
take into account the effects of the  spontaneous breaking of the GUT gauge
group to the SM gauge group. In consequence,  the three
SM gauge couplings are affected differently by these corrections, so that
after passing the super-heavy threshold they become equal and evolve as a
unique coupling towards the Planck scale.
Furthermore, the calculation of the associated decoupling constants is much more involved and very model
dependent. Another subtle point concerns the choice of gauge within the full
theory so that the gauge
invariance of the effective theory, obtained by integrating out the heavy
degrees of freedom, is also
maintained~\cite{Weinberg:1980wa}. 

Currently, the super-heavy threshold corrections to the gauge couplings are
known at the one-loop level for a general
model~\cite{Hall:1980kf,Weinberg:1980wa}. However, at the two-loop level there
are only the attempts of Refs.~\cite{Martens:2010pe, Martens:2011} towards a
general derivation. Precisely, this computation covers only theories without
super-heavy fermions and with a rather simplified structure for the GUT
breaking scalars. Also, trilinear interactions in the scalar potential are not
considered. Thus, their applicability to specific models is rather restricted.

\begin{figure}[t]
  \centering
  \includegraphics[width=1.\linewidth]{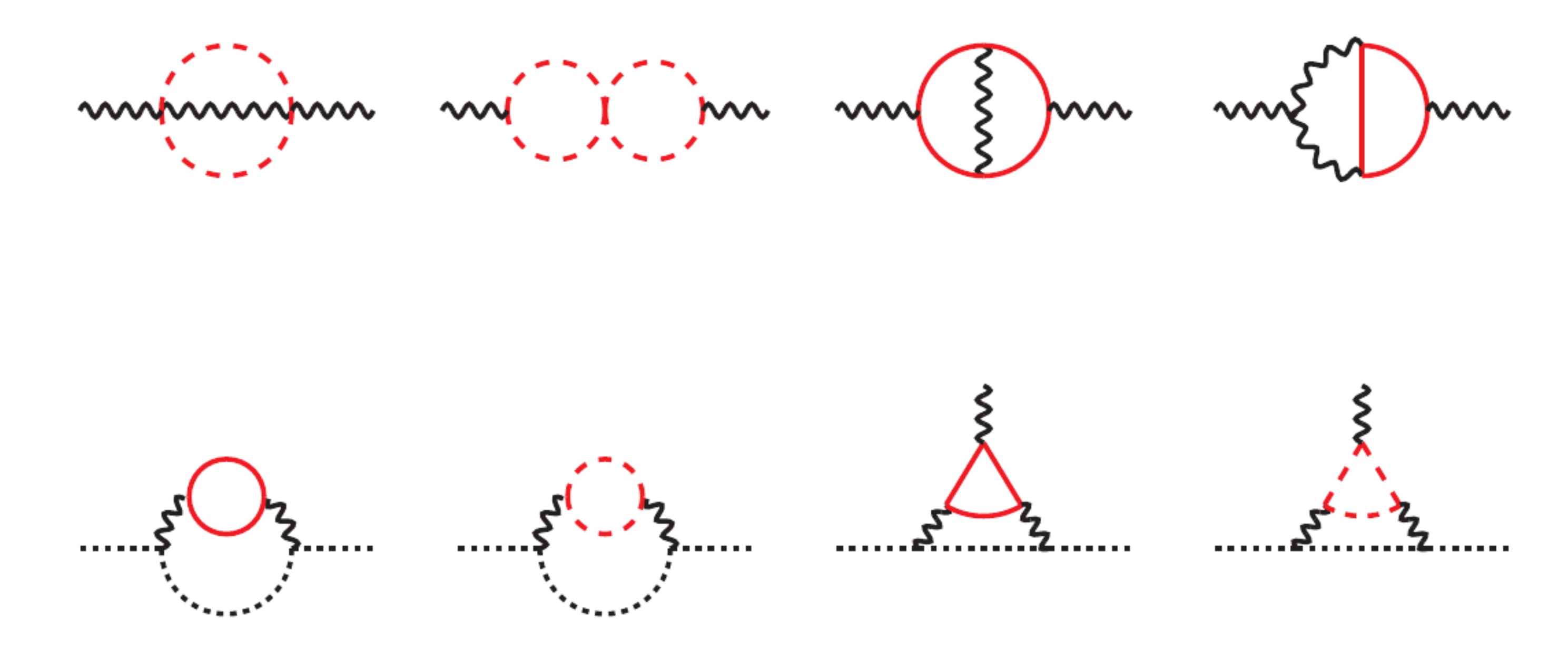}
  \caption[]{\label{fig:gutdia} Feynman diagrams
    contributing to the two-loop threshold corrections at the GUT scale. Bold
    (red) lines represent fields with mass of $\mathcal O(M_{\rm GUT})$ and
    thin lines massless fields.  Furthermore, curly lines denote gauge bosons,
    dotted lines ghosts, dashed lines scalar fields and solid lines fermions.}
\end{figure}

For exemplification, we show in Fig.~\ref{fig:gutdia} 
sample Feynman diagrams contributing at the
two-loop level to the decoupling constant of the gauge couplings. 
As usual, to compute the decoupling constants one considers
Green's functions involving only light external particles and at least one
heavy field within the loops. In Fig.~\ref{fig:gutdia} the contributions to
the self-energies of the light gauge bosons and their associated ghosts, and
to the gauge boson-ghost three-point function are shown.

A numerical analysis of the effect of the two-loop GUT threshold
corrections on gauge coupling unification was performed in
Ref.~\cite{Martens:2010pe}.  It turned out that, for models containing large
representations, the effects of the two-loop GUT threshold corrections exceed
by almost an order of magnitude those induced by the current experimental
uncertainties of the gauge coupling constants.  In consequence, the GUT
threshold corrections necessarily have to be taken into account in any
phenomenological study concerning the gauge coupling unification. An
impression of their numerical effects in the minimal SUSY SU(5) model can be
seen in Fig.~\ref{fig:gutunif}(b) (however, only at the one-loop level).

\subsection{\label{sub::LET}Application I: low-energy theorems}

There is a close connection between the decoupling constants and
the effective coupling of a Higgs boson to gluons and light quarks
which at first sight is quite surprising. It is established by the 
so-called low-energy theorem (LET) which relates the decoupling constants
to the Wilson coefficients in the effective Lagrangian
\begin{eqnarray}
  {\cal L}_{\rm eff} &=& -\frac{H}{v} 
  \left(C_1 \frac{1}{4} G_{\mu\nu} G^{\mu\nu}
    + C_2 m_b \bar{\psi}_b\psi_b
  \right)
  \,,
  \label{eq::leff2}
\end{eqnarray}
where the first part has already been shown in Eq.~(\ref{eq::leff}).
The term proportional to $C_2$ has been specified to the bottom quarks;
similar contributions also exist for the other light quarks.
In Ref.~\cite{Chetyrkin:1997un} the following LETs have been
derived
\begin{eqnarray}
  C_1 &=& - \frac{m_t \partial}{\partial m_t} \zeta_{\alpha_s} \,,
  \nonumber\\
  C_2 &=& 1 + \frac{m_t \partial}{\partial m_t} \zeta_{m_b}  \,,
  \label{eq::let}
\end{eqnarray}
where we refer to~\cite{Chetyrkin:1997un} for a proper definition of the
decoupling constants (with $\zeta_{\alpha_s} = (\zeta_{g})^2$).

At lowest order the LETs given in Eq.~(\ref{eq::let}) can easily be motivated.
Consider, e.g., $C_1$ which can be computed from the one-loop top quark
triangle with two gluons and a zero-momentum Higgs boson as external particle.
On the other hand, only the one-loop heavy-top diagram of the gluon propagator
contributes to $\zeta_{\alpha_s}$. Zero-momentum top
quark-Higgs boson coupling are obviously generated by taking derivatives with
respect to the top quark mass which leads to the first line in
Eq.~(\ref{eq::let}) expanded to one-loop order.  
At higher order this simple picture does not work any
more due to the fact that also other Green's functions enter
$\zeta_{\alpha_s}$, e.g., the ghost two-point function.

It is interesting to note that Eq.~(\ref{eq::let}) only contains logarithmic
derivatives. Thus, it is sufficient to know the $\ln(m_t^2)$ terms of
$\zeta_{\alpha_s}$ and $\zeta_{m_q}$ in order to obtain $C_1$ and $C_2$ at the
respective loop order. Since $m_t$ is the only dimensionful physical scale it
actually occurs in the combination $\ln(\mu^2/m_t^2)$ and thus the logarithms
at order $\alpha_s^{(n+1)}$ can be reconstructed from the order $\alpha_s^{n}$
with the help of renormalization group functions.  In
Refs.~\cite{Schroder:2005hy,Chetyrkin:2005ia} this has been exploited to
obtain the five-loop result for $C_1$ from the four-loop result of
$\zeta_{\alpha_s}$.\footnote{Note that the fermionic contribution of the
  five-loop beta function, which is not yet known, enters the five-loop term
  of $C_1$ as a free
  parameter.}
In beyond-SM theories this trick cannot be applied due to the presence of more
than one particle mass.

For earlier work on LETs we want to refer to
Refs.~\cite{Kilian:1995tra,Kniehl:1995tn} and for the application of LETs in
supersymmetric theories we want to mention the
works~\cite{Degrassi:2008zj,Mihaila:2010mp,Kurz:2012ff}.
In Ref.~\cite{Grozin:2011nk} it has been applied to QCD involving more than
one heavy quark.

\subsection{\label{sub::su5}Application II:  gauge coupling
  unification in models based on the SU(5) group}

The quantum numbers of the SM fermions together with the apparent convergence 
of the strong and electroweak couplings at energies below the Planck scale 
point towards a unified description of the SM interactions. 
Furthermore, one of the fundamental predictions of a GUT is the existence 
of baryon and lepton number violating interactions which can manifest themselves 
at low energy via matter instability  
(for a review see for example Ref.~\cite{Nath:2006ut}). 
Though the decay of the proton has not been observed so far, the lower bound on
the proton lifetime together with the low-energy values of the SM gauge couplings 
and the SM fermion masses and mixing provide us severe constraints
on the class of viable GUT models.

Gauge coupling unification is highly sensitive to the mass spectrum. This
property allows us to probe the unification assumption through precision
measurements of low-energy parameters like the gauge couplings at the
electroweak scale or the mass spectrum and precision calculations. Through
such analyses, we can make predictions for some of the mass parameters of the
models that have to be compared with the constraints derived from the
non-observation of the proton decay. In addition, GUTs might predict
interesting signatures at the LHC (see, e.g., Ref. \cite{ATLAS1} for a recent
search for heavy fermionic triplets and Ref.~\cite{Khachatryan:2014ura} for
searches for scalar leptoquarks). These additional constraints from the LHC
can, in some cases, be sufficient to even rule out
models~\cite{Bajc:2007zf,Franceschini:2008pz,delAguila:2008cj,Chang:2012ta,DiLuzio:2013dda}.

In the following, we restrict our discussion to GUTs based on the SU(5) gauge
group. Although phenomenologically there are other theories which are better
motivated, SU(5) GUTs are most predictive. For example, one of the few
absolute certainties about grand unification today is that the original SU(5)
model of Georgi and Glashow (GG) \cite{Georgi:1974sy} is ruled out. In
particular, the failure of the minimal model can be attributed partially to
the lack of gauge coupling unification
\cite{Ellis:1990wk,Langacker:1991an,Amaldi:1991cn}.

Let us briefly recall the reason why gauge coupling unification fails within
the minimal GG model.\footnote{For the definition of $\alpha_1$, $\alpha_2$
  and $\alpha_3$ see Eq.~(\ref{eq::alpha_123}).} While $\alpha_2$ and $\alpha_3$ meet around $10^{16}$
GeV, $\alpha_1$ and $\alpha_2$ intersect already at
about $10^{13}$~GeV, at odds with the bounds enforced by the nonobservation of
the proton decay. 
More precisely, model independent upper bounds on the proton
lifetime~\cite{Dorsner:2004xa} together with the latest experimental 
data from the Super-Kamiokande observatory~\cite{Nishino:2012ipa} 
imply a conservative lower bound on the unification scale $M_G$ of about
$10^{15.5} \ \rm{GeV}$.
Hence, the
key ingredients for a viable unification pattern are additional particles
charged under the ${\rm SU(2)}_L$ group that delay the meeting of $\alpha_1$
and $\alpha_2$.  This is, essentially, the philosophy behind two recent
proposals where an extra scalar representation $15_H$
\cite{Dorsner:2005fq,Dorsner:2005ii}, or alternatively, a fermionic
representation $24_F$ \cite{Bajc:2006ia,Bajc:2007zf} are added to the field
content of the model.  In both cases, the extra degrees of freedom have the
correct quantum numbers to restore unification by properly modifying the running
of the gauge couplings.

\begin{figure}[t]
  \begin{center}
      \includegraphics[angle=0,width=\linewidth]{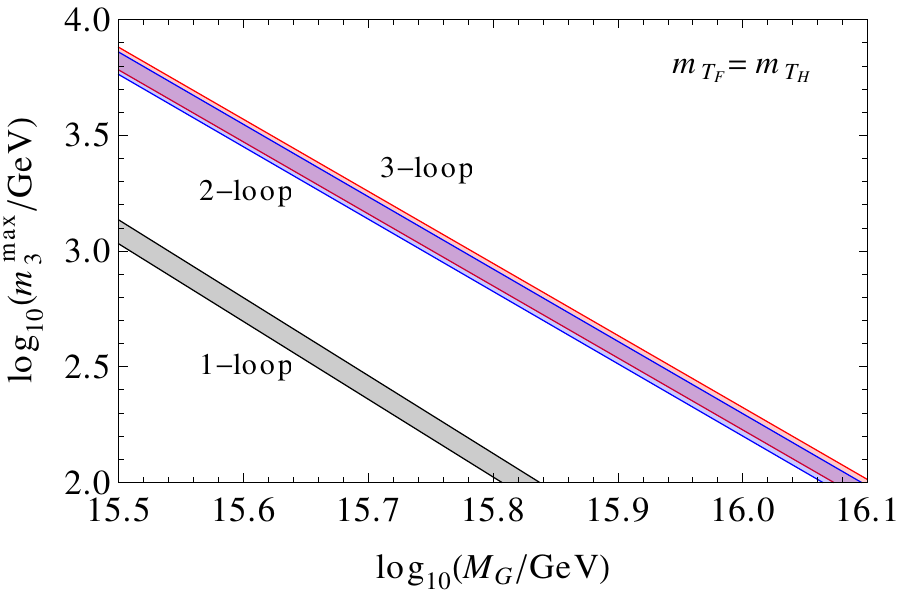}
      \caption[]{\label{fig:m3gut} The maximal value of the effective
        triplet mass $m_3$ in the SU(5)+$24_F$ model as a function of the
        unification scale $M_G$.  The black, blue and red bands (from
        bottom-left to top-right) correspond respectively to the one-, two-
        and three-loop running analysis.  The error bands are due to the
        $1\sigma$ uncertainties on the low-energy electroweak couplings
        $\alpha_1 (M_Z)$ and $\alpha_2 (M_Z)$ (Figure taken from
        Ref.~\cite{DiLuzio:2013dda}).}
  \end{center}
\end{figure}

In particular, the study performed in Ref.~\cite{DiLuzio:2013dda} proved that
a three-loop analysis of the gauge coupling unification is required within the
SU(5)+$24_F$ model. As can be read from the Fig.~\ref{fig:m3gut}, the two-loop
corrections to the mass of the electroweak triplets $m_3$ for a fixed
unification scale $M_G$ is of the same order of magnitude as the one-loop
contributions and amount to several TeV.  The three-loop corrections are
rather small (hundreds of GeV) and lie within the uncertainty band of the
two-loop results.  Thus, the three-loop order analysis is required in order to
reduce the theoretical uncertainties at the level of parametric uncertainties
induced by the low-energy measurements of the electroweak gauge
couplings. Moreover, the achieved theoretical accuracy together with
experimental data in the reach of the LHC and the Super-Kamiokande observatory
will provide us with sufficient information to even disprove the model.

Another example of a predictive GUT model is the minimal SUSY
SU(5)~\cite{Dimopoulos:1981zb,Sakai:1981gr}. It has the important feature that
one can derive unambiguous correlations among its parameters and even rule out
the model once sufficiently precise experimental data become
available. Immediately after the formulation of the minimal SUSY SU(5) it has
been noticed that within SUSY GUTs new dimension-five operators cause a rapid
proton decay~\cite{Sakai:1981pk,Weinberg:1981wj}.  This aspect was intensively
studied over the last thirty years with the extreme conclusions of
Refs.~\cite{Goto:1998qg,Murayama:2001ur} that the minimal SUSY SU(5) model is
ruled out by the combined constraints from proton decay and gauge coupling
unification. However, careful analyses have shown that taking into account
fermion mixing~\cite{Bajc:2002bv} or higher dimensional operators induced at
the Planck scale~\cite{EmmanuelCosta:2003pu,Bajc:2002pg}, one can
substantially weaken the constraints and the minimal model remains a valid
theory. Also the high-precision analysis of Ref.~\cite{Martens:2010nm}
confirmed that the minimal SUSY SU(5) model cannot be excluded by the current
experimental data. In particular, one observes an increase of the super-heavy
Higgs triplet mass by about an order of magnitude when three-loop 
effects are considered.  These results attenuate substantially the tension
between the theoretical predictions and the constraints derived from the
experimental data.

Beyond the minimal version of the SUSY SU(5) model, the interplay between the
theoretical predictions and the experimental data is not completely
determined. For example, the most popular extension of the minimal SUSY SU(5)
model, the Missing Doublet Model~\cite{Masiero:1982fe,Grinstein:1982um},
designed to avoid unnatural doublet-triplet splitting, cannot be excluded
using only the currently available theoretical and experimental data. The
model contains additional free parameters as compared to the minimal model
and, consequently, is affected by large theoretical uncertainties, so that no
firm conclusion can be drawn.

\begin{figure}[ht]
  \begin{center}
    \begin{tabular}{c}
      \includegraphics[width=\linewidth]{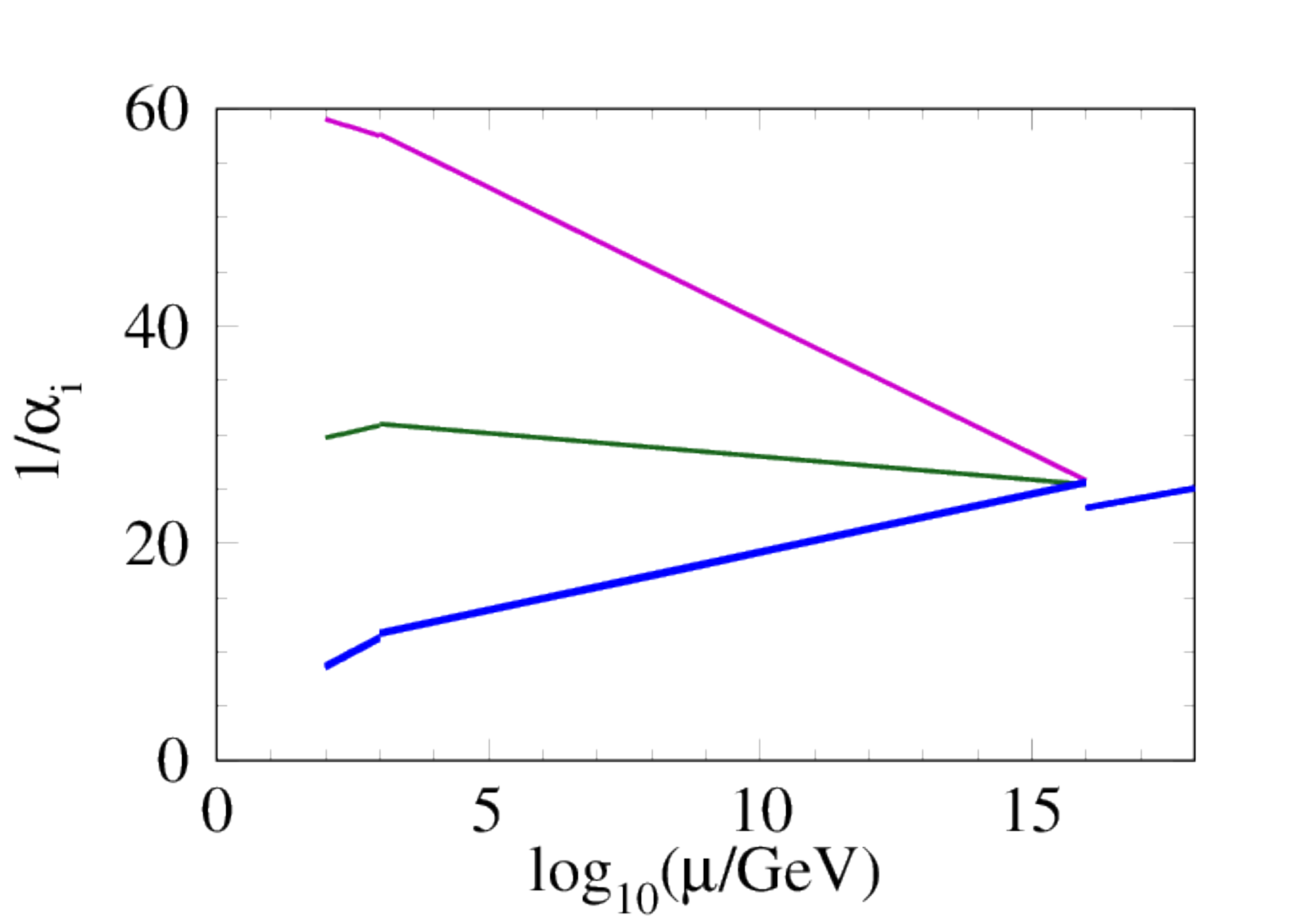}
      \\ (a)\\
      \includegraphics[width=\linewidth]{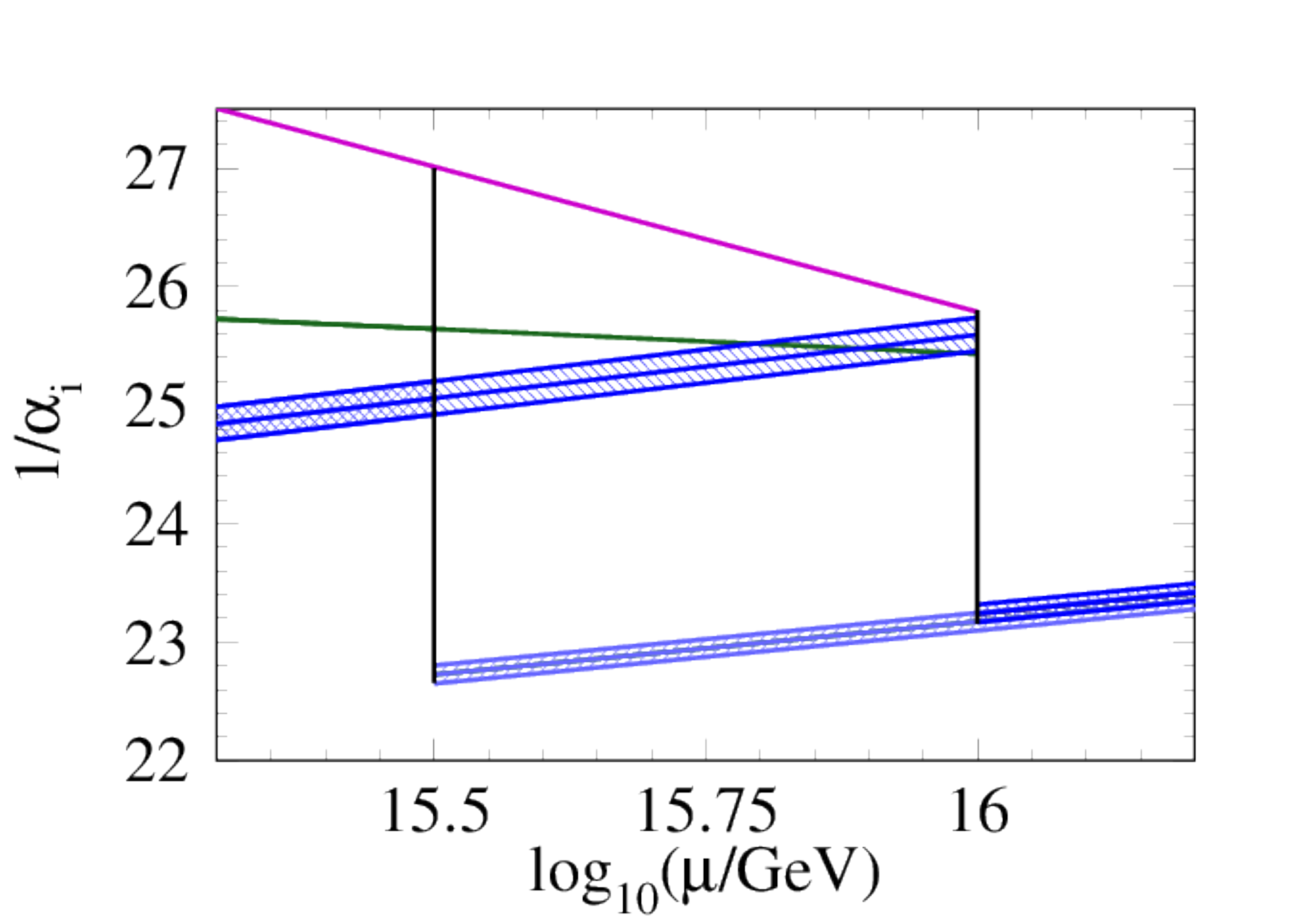}
      \\ (b)
    \end{tabular}      
    \caption[]{\label{fig:gutunif} Running of the gauge couplings from the
      electroweak to the Planck scale from Ref.~\cite{Martens:2010nm}. The
      discontinuity for $\mu=\mususy=10^3$~GeV and $\mu=\mugut=10^{15.5}$~GeV
      are clearly visible. In panel (b) an enlargement of (a) for the region
      around $\mu=\mugut$ is shown where for the decoupling the two values
      $\mugut=10^{16}~\mbox{GeV}\approx 3.2\cdot 10^{15}$~GeV and
      $\mugut=10^{16}$~GeV have been chosen. Figure taken from
      Ref.~\cite{Martens:2010nm}. } 
  \end{center}
\end{figure}

For illustration, we briefly discuss the analysis of the gauge coupling
unification in the SUSY SU(5) model at three-loop level. Crucial input
parameters for this analysis are the precise values of the gauge couplings at
the electroweak scale. For their determination from the experimentally
measured observables, one has to take into consideration threshold corrections
at the Z-boson mass and at the top quark mass (for a detailed description see
Ref.~\cite{Martens:2010nm}). Furthermore, one applies the ``running and
decoupling'' approach described in the previous sections. The dependence on
the heavy particle masses becomes explicit through the decoupling
constants. The constraint of gauge coupling unification translates into
restrictions (usually expressed as correlations) on the mass spectrum.

In Fig.~\ref{fig:gutunif} the evolution of the SM gauge couplings from
the electroweak up to the Planck scale is shown in the minimal SUSY SU(5)
model. The specific threshold corrections for this model are those due to
supersymmetric particles at the TeV scale and those due to super-heavy
particles at around $10^{16}$~GeV.  The associated decoupling scales have been
chosen at $\mususy = 1000$~GeV and $\mugut = 10^{16}$~GeV. One can clearly see
the discontinuities at the matching scales and the change of the slopes when
passing them. The (blue) band around the strong coupling corresponds to the
present experimental uncertainty.  In panel (b) the region around $\mu =
10^{16}$~GeV is enlarged which allows for a closer look at the unification
region. Here, the decoupling of the super-heavy particles is performed at two
different values of $\mugut$. One observes quite different threshold
corrections leading to a nice agreement of the (common) gauge coupling above
$10^{16}$~GeV. On the one hand, the plot proves that the GUT threshold
corrections are essential for a successful unification. On the other hand, the
dependence on $\mugut$ is interpreted as a measure of theoretical
uncertainties because this parameter is not fixed by the theory. Thus, the
three-loop analysis is sufficient to cope with the current experimental
precision.

To summarise, threshold corrections at the GUT scale are indispensable
ingredients for precision analyses of gauge coupling unification. In
particular, they have the virtue to provide us with the necessary information
about the specific model dependence. In this way, we are able to perform
consistency tests of various GUTs and, in some cases, even to refute models.

\subsection{\label{sub::Hpot}Application III: stability of  the SM vacuum}

With the discovery of a Higgs boson at the LHC, the question of the SM vacuum
stability received a renewed attention. This is because the mass of the Higgs
boson in the SM is an important indicator for the presence of new physics at
high energy scales. It is well known that, if the Higg boson mass satisfies
the condition $M^{\rm meta}<M_h<M^{\rm Landau}$ then the SM is a consistent
theory from the electroweak scale up to the Planck scale. The upper limit
originates from the requirement that the Higgs self-coupling remains in
the perturbative regime up to the Planck scale. Explicitly, the occurrence
of the Landau pole is avoided. The estimated value for $M^{\rm Landau}$ is
around $175$~GeV~\cite{Hambye:1996wb}, which is excluded by the direct
searches at the LHC and the Tevatron. The lower limit is derived from
instability constraints. The existence of a Higgs boson with a mass smaller
than $M^{\rm meta}$ would imply that new physics below the Planck scale is
required to stabilize the SM vacuum. The numerical value for $M^{\rm meta}$
was estimated to $111$~GeV~\cite{Espinosa:2007qp}. In consequence, for a Higgs
boson with mass $M_h\approx125$~GeV one distinguishes two situations:
either $M_h>M^{\rm stability}$ and the electroweak vacuum is absolutely
stable, or $M^{\rm meta}<M_h<M^{\rm stability}$, which corresponds to a
metastable vaccum with a life-time exceeding that of the Universe.

The precise determination of $M^{\rm stability}$ was the subject of numerous
recent
analyses~\cite{Bezrukov:2012sa,Degrassi:2012ry,Alekhin:2012py,Buttazzo:2013uya}.
The absolute stability bound on the Higgs mass $M^{\rm stability}$ is defined
as the value for which the effective potential at the electroweak minimum,
$\phi_{\rm{ew}}$, and at a second minimum at large field values,
$\tilde{\phi} > \phi_{\rm{ew}}$, are the same. The equations to be solved read
\begin{eqnarray}
\label{effpot}
 V(\phi_{\rm{ew}}, M^{\rm  stability}) &=& V(\tilde{\phi}, M^{\rm  stability})  \, , \nonumber\\  
 \left. \frac{\partial V}{\partial \phi} \right|_{\phi_{\rm{ew}}}
 &=& \left. \frac{\partial V}{\partial \phi} \right|_{\tilde{\phi}}
 = 0 \, .  
\end{eqnarray}
Although the effective potential as well as the positions of its extrema are
both gauge dependent, the solution for $M^{\rm stability}$ is gauge
independent as shown in Ref.~\cite{DiLuzio:2014bua}.

\begin{figure}[t]
  \begin{center}
      \includegraphics[angle=0,width=\linewidth]{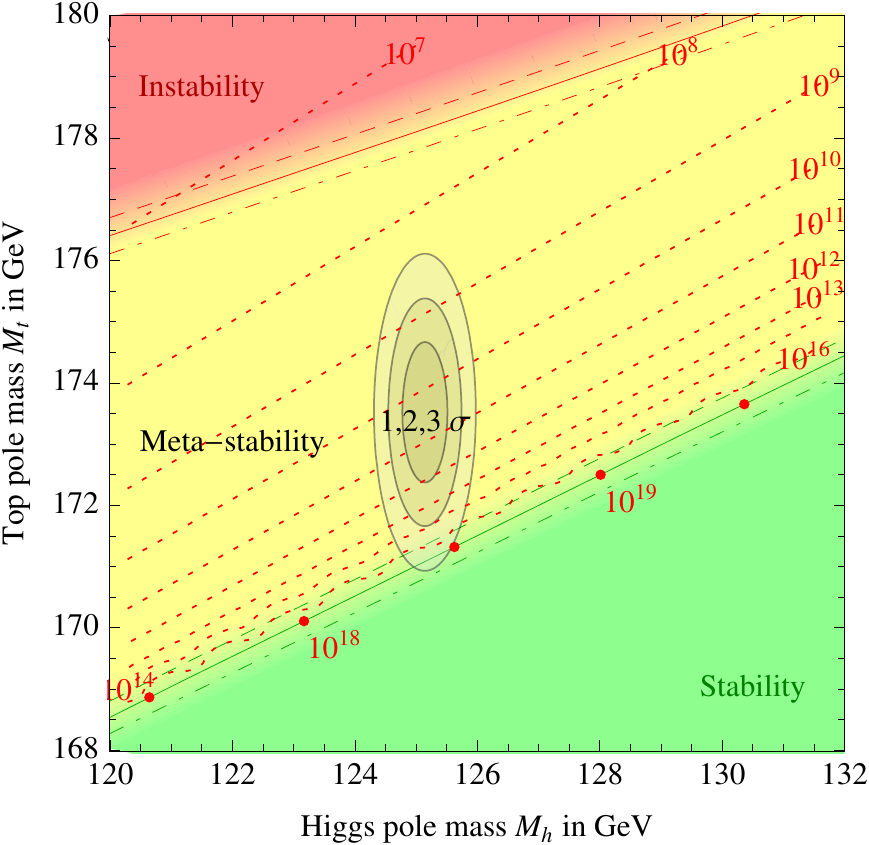}
      \caption[]{\label{fig:smstab} SM phase diagram in terms of the Higgs
        boson $M_h$ and top quark $M_t$ masses. The dotted contour-lines show
        the instability scale $\Lambda_{\rm SM}$ in GeV for
        $\alpha_s(M_Z)=0.1184$. Figure taken from
        Ref.~\cite{Buttazzo:2013uya}.}
  \end{center}
\end{figure}

The most advanced recent works use for the calculation of $M^{\rm stability}$
the two-loop SM effective potential~\cite{Ford:1992pn,Martin:2001vx} and the
three-loop RGEs for the SM
couplings~\cite{Mihaila:2012fm,Mihaila:2012pz,Bednyakov:2012rb,Chetyrkin:2012rz,Bednyakov:2012en,Chetyrkin:2013wya,Bednyakov:2013eba},
necessary to resum the large logarithms that occur in the effective potential
$V$ for large field values. In addition, two-loop corrections to the relations
between the \msbar{} parameters and the physical observables have been
considered in Refs.~\cite{Bezrukov:2012sa,Degrassi:2012ry,Buttazzo:2013uya}.
The phenomenological implications of the new determination of $M^{\rm
  stability}$ are summarised in the phase diagram shown in
Fig.~\ref{fig:smstab}. The regions of
stability, metastability and instability of the SM vacuum are shown for Higgs
boson ($M_h$) and top quark ($M_t$) masses in the range corresponding to the
measured values. Also shown are contour lines indicating the instability scale
$\Lambda_{\rm SM}$, defined as the scale where new physics is required to
stabilize the SM effective potential. It appears that the measured values for
$M_h$ and $M_t$ are rather special, in the sense that, the SM vacuum lies at
the border between stability and metastability. One also concludes that a
metastable electroweak vacuum can comply with the data and new physics below
the Planck scale is not necessarily implied.

\begin{figure}[t]
  \begin{center}
      \includegraphics[angle=0,width=\linewidth]{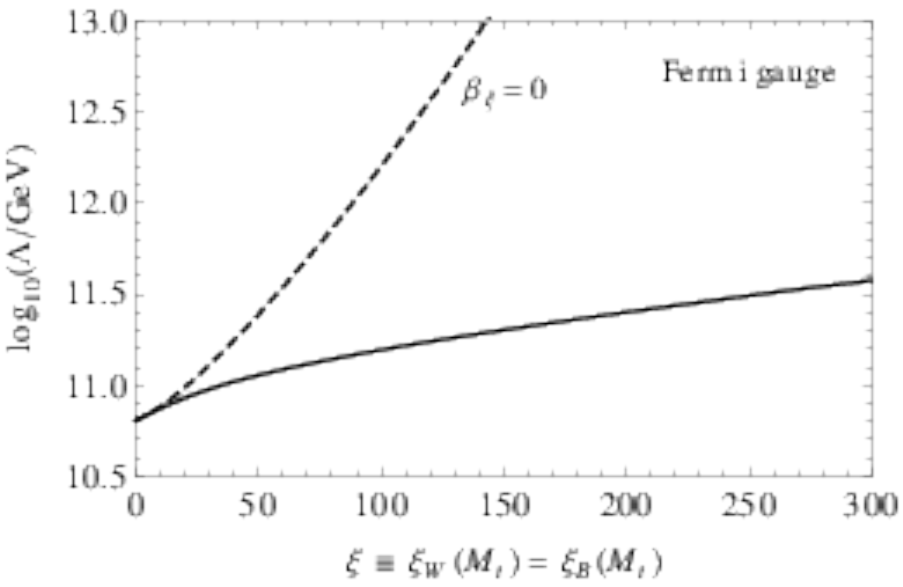}
      \caption{\label{fig:sminst} Instability scale as a function of $\xi
        \equiv \xi_W (M_t) = \xi_B (M_t)$ for the Fermi gauge.  The dashed
        curve corresponds to the case where the gauge fixing parameters do
        not run.}
  \end{center}
\end{figure}

Note that the interpretation of the SM instability scale as the  scale
$\Lambda$, where 
\begin{eqnarray}
  V(\phi)&>&V(\phi_{\rm{ew}})\quad\mbox{for all}\quad \phi<\Lambda 
  \label{eq:cutoff}
\end{eqnarray}
is problematic.  The effective potential is gauge dependent, and it renders
$\Lambda$ as solution of the inequality~(\ref{eq:cutoff}) gauge dependent as
well. The numerical effects due to gauge dependence are rather large as has
been explicitly shown in
Refs.~\cite{DiLuzio:2014bua,Nielsen:2014spa,Andreassen:2014gha}. The
instability scale can vary by few orders of magnitude in $R_\xi$-like gauges,
depending on the choice of the gauge fixing parameters. This behaviour is
illustrated in Fig.~\ref{fig:sminst} taken from
Ref.~\cite{DiLuzio:2014bua}. Here, the dependence of the SM instability scale
[defined as in inequality~(\ref{eq:cutoff})] on a common gauge fixing
parameter in the electroweak sector is displayed for the special choice of a
Fermi gauge (see Ref.~\cite{DiLuzio:2014bua} for precise definition).  As can
be read from the figure, the resummation obtained through the use of running
gauge fixing parameters is essential to reduce the variation of the
instability scale. However, even the RGE improved prediction for
$\Lambda_{\rm SM}$ varies by an order of magnitude for gauge fixing parameters
in the perturbative regime. Let us at this point stress that
significant variations of the instability scale are also observed
when changing the form of the gauge fixing term in the Lagrange density.

To conclude, a metastable electroweak vacuum can comply with the present
experimental data, and new physics below the Planck scale is not necessarily
required. Furthermore, the physical threshold for new physics can be
determined only after specifying an ultraviolet completion of the SM. Its
determination from the stability requirement of the SM potential is
theoretically inconsistent due to the gauge dependence as discussed above.



\section{\label{sec::sum}Summary}

With the discovery of a Higgs boson particle at the LHC physics has entered a
new era which is at the moment dominated by detailed and careful studies of
its properties.  The central question about the particle discovered at about
125~GeV is whether this is ``the Higgs boson" of the SM or only one degree of
freedom of a bigger theory. The fundamental difficulty to identify new physics
signatures at the LHC, so far, initiated numerous precision analyses focused
on the Higgs sector.

In this review we report on recent precision calculations and computational
developments.  In particular, the following topics are highlighted: the
determination of the Higgs boson mass within the MSSM at three-loop accuracy;
the calculation of the Higgs production in the SM and the MSSM through
N$^3$LO and NNLO, respectively; the computation of
renormalization group functions within the SM and its extensions at the
three-loop order. Special emphasis was put on the realisation of the gauge
coupling unification within BSM theories and on the study of the SM vacuum
stability.

Hopefully, accurate understanding of the Higgs phenomenology obtained through
high precision analyses together with new information from experiments at the
LHC will provide us a tool for exploring new physics and to find explanations
for some of the long-standing questions in particle physics.






\section*{Acknowledgements}
This work is supported by the Deutsche
Forschungsgemeinschaft in the Sonderforschungsbereich Transregio~9
``Computational Particle Physics''.
We would like to thanks Maik H\"oschele, Jens Hoff and Sven Moch for
comments to the manuscript.







\end{document}